\documentclass[fleqn,usenatbib]{mnras}

\usepackage{newtxtext,newtxmath}
\usepackage[T1]{fontenc}
\usepackage{ae,aecompl}

\usepackage{graphicx}	% Including figure files
\usepackage{amsmath}	% Advanced maths commands
\usepackage{amssymb}	% Extra maths symbols
\usepackage{dblfloatfix}

\newcommand{\galform}{\textsc{galform}}
\newcommand{\coco}{\textsc{coco}}
\newcommand{\Color}{\textsc{color}}
\newcommand{\PGadget}{\textsc{P-Gadget-3}}

\newcommand{\Subfind}{\textsc{subfind}}

\title[The abundance of ultrafaint satellites]{The little things matter: relating the abundance of ultrafaint satellites to the hosts' assembly history}

\author[S. Bose et al.]{
Sownak Bose,$^{1}$\thanks{E-mail: sownak.bose@cfa.harvard.edu}
Alis J. Deason,$^{2}$
Vasily Belokurov,$^{3}$
and Carlos S. Frenk$^{2}$
\\
$^{1}$Center for Astrophysics | Harvard \& Smithsonian, 60 Garden St., Cambridge, MA 02138, USA\\
$^{2}$Institute for Computational Cosmology, Durham University, South Road, Durham, DH1 3LE, UK\\
$^{3}$Institute of Astronomy, University of Cambridge, Madingley Road, Cambridge, CB3 0HA, UK
}

\date{Accepted XXX. Received YYY; in original form ZZZ}

\pubyear{2019}

\begin{document}
\label{firstpage}
\pagerange{\pageref{firstpage}--\pageref{lastpage}}
\maketitle

\begin{abstract}
  Ultrafaint dwarf galaxies ($M_\star\lesssim10^{5-6}\,{\rm M}_\odot$)
  are relics of an early phase of galaxy formation. They contain some
  of the oldest and most metal-poor stars in the Universe which likely
  formed before the epoch of hydrogen reionisation. These galaxies are
  so faint that they can only be detected as satellites of the Milky
  Way. They are so small that they are just barely resolved in current
  cosmological hydrodynamics simulations. Here we combine very high
  resolution cosmological $N$-body simulations with a semi-analytic
  model of galaxy formation to study the demographics and spatial
  distribution of ultrafaint satellites in Milky Way-mass haloes. We
  show that the abundance of these galaxies is correlated with the
  assembly history of the host halo: at fixed mass, haloes assembled
  earlier contain, on average, more ultrafaint satellites today than
  haloes assembled later. We identify simulated galactic haloes that
  experience an ancient {\it Gaia}-Enceladus-Sausage-like and a recent
  LMC-like accretion event and find that the former occurs in 33\% of
  the sample and the latter in 9\%. Only 3\% experience both events
  and these are especially rich in ultrafaint satellites, most
  acquired during the ancient accretion event. Our models predict that
  the radial distribution of satellites is more centrally concentrated
  in early-forming haloes. Accounting for the depletion of satellites
  by tidal interactions with the central disc, we find a very good
  match to the observed radial distribution of satellites in the Milky
  Way over the entire radial range. This agreement is mainly due to
  the ability of our model to track `orphan' galaxies after their
  subhaloes fall below the resolution limit of the simulation.
\end{abstract}

\begin{keywords}
Galaxy: formation -- galaxies: dwarf -- {\it (galaxies:)} Local Group -- {\it (cosmology:) reionization} -- methods: numerical
\end{keywords}

%%%%%%%%%%%%%%%%%%%%%%%%%%%%%%%%%%%%%%%%%%%%%%%%%%

%%%%%%%%%%%%%%%%% BODY OF PAPER %%%%%%%%%%%%%%%%%%

\section{Introduction}
\label{sect:intro}

Dark matter haloes are the fundamental units of cosmic structure. They are the sites where galaxies form and the details of their assembly directly influence the properties of the galaxies within them. In the $\Lambda$CDM model dark matter haloes grow hierarchically through the continuous accretion of mass in the form of discrete lumps and diffuse matter \citep[e.g.][]{Frenk1985,Wang2011,Wechsler2002,McBride2009,Fakhouri2010,Correa2015}. The faint population of dwarf galaxies that dominate in number are amongst the oldest objects in the cosmos while, at the other end of the mass range, rare, rich clusters of galaxies assemble later and are still growing at the present day.

Much of what we have learnt about the growth of dark haloes has come from high-resolution numerical simulations. Inferring the assembly of haloes from observational data is considerably more challenging and can only be done by means of indirect tracers. Unsurprisingly, this enterprise has been most successful in the case of our own galaxy, the Milky Way, for which the data are now more complete and comprehensive than ever before. Through the combination of a variety of probes including the kinematics of individual stars, globular clusters and satellite galaxies, increasingly tight constraints on the mass of the Galaxy's halo -- one of the most important parameters in astrophysics -- have been obtained \citep[e.g.][]{Deason2012,BoylanKolchin2013,Patel2018,Callingham2019,Watkins2019}. Furthermore, the measurement of high precision stellar proper motions by the {\it Gaia} satellite \citep{Gaia2018} has improved our understanding of the assembly of the Galaxy manyfold.

There is now increasing evidence that the formation of the Milky Way was punctuated by two distinct events, separated in time by several billion years. The most familiar is the accretion of the Small and Large Magellanic Clouds (SMC and LMC), which are thought to have fallen into the potential of the Milky Way's halo 2-3 Gyrs ago and to be on their first orbit in the Galactic potential \citep[e.g.][]{Kallivayalil2006,Besla2007,Sales2011,BoylanKolchin2011,Besla2015,Shao2018}. More recently, galactic archaeology based on metal-rich halo stars in {\it Gaia} data has provided evidence for an ancient dwarf galaxy merger, similar in mass to the LMC, between 8 and 11 Gyrs ago, roughly around the time of the formation of the Galactic disc
\citep{Belokurov2018,Helmi2018,Myeong2018}. The merger remnant, known as the {\it Gaia}-Enceladus or {\it Gaia} ``Sausage''-- so-named for the elongated appearance of its constituent stars in velocity space -- provides a pointer to the early accretion history of the Galactic halo. Numerical simulations are needed to determine the frequency of events of this kind and thus to quantify how typical the Milky Way is compared to the overall population of galaxies \citep[e.g.][]{Fattahi2019, Mackereth2019}.

Galaxy mergers are important for more than just adding mass to, and disturbing, the central galaxy: they also bring in large populations of satellite galaxies that may survive until the present day \citep[e.g.][]{Deason2015,Jethwa2016,Sales2017,Dooley2017,Shao2018,Jahn2019}. A complete census of the satellite population of galactic haloes offers a powerful constraint on physical processes associated with galaxy formation \citep[e.g.][]{Bullock2000,Benson2002a,Bovill2009,Font2011,Brooks2014,Sawala2016,Wetzel2016,Munshi2019}, as well as on the nature of the dark matter itself \citep[e.g.][]{Maccio2010b,Lovell2012,Kennedy2014,Lovell2016,Newton2018}. In both instances, the crucial population are the `ultrafaint' dwarf galaxies ($M_\star\lesssim10^{5-6}\,{\rm M}_\odot$) whose detection, whilst still notoriously difficult, has improved greatly with the advent of the Sloan Digital Sky Survey \citep[SDSS,][]{Adelman2007,Alam2015}, the Dark Energy Survey \citep[DES,][]{Bechtol2015,Drlica2015,Kim2015,Koposov2015} and Pan-STARRS1 \citep{Laevens2015,Chambers2016}.

Even at fixed mass, observations find a large degree of scatter in the abundance of satellites surrounding galaxies similar to our own \citep[e.g.][]{Zaritsky1993,Guo2011,Wang2012,Geha2017,Kondapally2018,Bennet2019}. The galactic populations we observe today may retain memory of the assembly histories of their hosts. Our aim in this work is to unify these two themes -- the assembly history of dark matter haloes and the population of ultrafaint satellites hosted within them -- into a single narrative.

The stellar masses of the ultrafaint satellites are so small ($M_\star \sim10^2 - 10^{5.5}\,{\rm M}_\odot$) that they are beyond the reach of the current generation of cosmological hydrodynamical simulations, such as {\sc Apostle} \citep{Sawala2016,Fattahi2016}, {\sc Auriga} \citep{Grand2017} or {\sc Fire} \citep{Hopkins2014,Hopkins2018}, which, at best, have a stellar mass resolution of $M_\star \sim10^3-10^4 \,{\rm M}_\odot$, although higher resolution may be achieved in simulations of {\it isolated} dwarf galaxies \citep[e.g.][]{Wheeler2018}. Instead, we use the technique of semi-analytic galaxy modelling in which mass resolution is not an issue. We graft our semi-analytic model, \galform{} \citep{Cole2000,Lacey2016}, onto merger trees constructed from the \coco{} $\Lambda$CDM dark matter simulations \citep{Hellwing2016,Bose2016}. This introduces a resolution scale in the {\it dark matter} of $\sim10^6\,{\rm M}_\odot$, which is more than adequate to resolve the haloes of the ultrafaint satellite population. With these simulations we investigate the diversity of formation histories of galaxies like the Milky Way and explore how the specific accretion events in the Milky Way may have shaped its present-day satellite content. The combination of high-resolution cosmological simulations with observational data from {\it Gaia} makes this a particularly timely endeavour.

This paper is organised as follows. In Section~\ref{sect:numerical} we describe the simulations and the semi-analytic model of galaxy formation used in this work. Our main results are presented in Section~\ref{sect:results}, in which we connect the populations of ultrafaint satellites of Milky Way-mass haloes to their assembly histories and predict the radial distribution of these satellites within their hosts. Finally, Section~\ref{sect:conclusions} provides a summary.

\section{Modelling techniques}
\label{sect:numerical}

This section provides an overview of the numerical setup used in this work. We describe the $N$-body simulations (Section~\ref{sect:sims}) that form the backbone of the semi-analytic model of galaxy formation, \galform{}, which is used to populate dark matter haloes with galaxies (Section~\ref{sect:Semia}).

\subsection{Simulation suite}
\label{sect:sims}

The $N$-body simulations we analyse are part of the {\it Copernicus Complexio} (\coco{}) suite of simulations
\citep{Hellwing2016,Bose2016}, a set of cosmological zoom-in $N$-body simulations that follow the evolution of over 12 billion high-resolution dark matter particles, each of mass $m_p = 1.6 \times 10^5\,{\rm M}_\odot$ from $\Lambda$CDM initial conditions. The zoom region, which is roughly 24 Mpc in radius, was extracted from a (100 Mpc)$^3$ periodic volume, the {\it Copernicus complexio Low Resolution} (\Color{}) simulation, in which the mass of each dark matter particle is $m_p = 8.8 \times 10^6\,{\rm M}_\odot$. The simulations were evolved from $z=127$ to the present day using the \PGadget{} code \citep{Springel2001,Springel2005}. Both \coco{} and \Color{} assume cosmological parameters derived from WMAP-7 data \citep{Komatsu2011}: $\Omega_m = 0.272$, $\Omega_\Lambda = 0.728$ and $h = 0.704$, where $h$ is related to the present-day Hubble constant, $H_0$, by $h = H_0/100{\rm kms}^{-1}{\rm Mpc}^{-1}$. The spectral index of the primordial power spectrum is $n_s = 0.967$, and the linear power spectrum is normalised at $z=0$ taking $\sigma_8 = 0.81$.

Dark matter haloes were identified using the friends-of-friends algorithm \citep{Davis1985}, which connects dark matter particles separated by at most 20 per cent of the mean interparticle separation in each volume. Gravitationally-bound substructures within each group are determined using the \Subfind{} algorithm \citep{Springel2001b}. To be included in the final halo catalogue, a subhalo is required to contain at least 20 bound dark matter particles, corresponding to a total mass of $3.2\times10^6\,{\rm M}_\odot$ in \coco{} and $1.8\times10^8\,{\rm M}_\odot$ in \Color{}. In what follows, the physical extent of a dark matter halo is defined by $r_{200}$, the radius within which the mean density of the halo is 200 times the critical density of the Universe. The mass of haloes is quoted in terms of $M_{200}$, the total mass in dark matter particles enclosed within $r_{200}$.

Substructures detected by \Subfind{} serve as the roots for building merger trees. We establish associations between subhaloes in subsequent output times by identifying objects that share some fraction of their most-bound particles between snapshots; the method is described in detail in \cite{Jiang2014}. The (sub)halo merger trees are then traversed to generate galaxy populations using the \galform{} semi-analytic model of galaxy formation, which we now describe.

\subsection{Semi-analytic galaxy formation}
\label{sect:Semia}

Semi-analytic models of galaxy formation provide a flexible and computationally efficient method for generating synthetic galaxy populations. Key advantages of these models over hydrodynamical simulations are their superior resolution and the relative ease with which it is possible to explore the vast parameter space describing the physics of galaxy formation. Semi-analytic modelling is, by now, a mature field and there exist several such models \citep[e.g.][]{Kauffmann1993,Cole1994,Somerville1999,Cole2000,Croton2006,Benson2012,Henriques2015}, which differ in the manner in which specific aspects of the galaxy formation process are treated. The general philosophy adopted in most of these models is similar: they follow the properties of (sub)haloes in merger trees and populate them with galaxies by solving sets of coupled differential equations that govern the cooling of gas in haloes, star formation, feedback from stars and black holes, chemical enrichment and the evolution of stellar populations. The free parameters describing these physical processes are calibrated by requiring that the model should reproduce a small selection of properties of the observed local galaxy population (typically at $z=0$).

We make use of the Durham semi-analytic model of galaxy formation, \galform{}, first presented in \cite{Cole1994} and refined in \cite{Cole2000}. In particular, we use the model developed in \cite{Lacey2016}, which unifies several features from previous versions, such as the assumption of a top-heavy initial mass function (IMF) in starbursts, which is required to reproduce the abundance of star-forming sub-millimetre galaxies \citep{Baugh2005}; the model of AGN feedback introduced by \cite{Bower2006}, which regulates the growth of massive galaxies; and a star formation law that depends on the molecular gas content of the interstellar medium \citep{Lagos2011}. Broad-band luminosities for the simulated galaxies are computed from the \cite{Maraston2005} stellar population synthesis model. The free parameters in the \cite{Lacey2016} model are calibrated so as to reproduce, at $z=0$, the optical and near-infrared luminosity functions, the black hole-bulge mass relation, the HI mass function and the fraction of early- and late-type galaxies. For a complete list of model parameters and their calibration we refer the reader to Section~4.2 of \cite{Lacey2016}.

Reionisation in the early Universe plays a decisive role in the
formation of the earliest galaxies. Ionising UV radiation from the first stars raises the temperature of the ambient hydrogen to $\sim10^4$K; this inhibits gas cooling in dark matter haloes whose effective virial temperature, $T_{{\rm vir}} \lesssim 10^4$K, halting their growth \citep[e.g.][]{Doroshkevich1967,Couchman1986,Rees1986,Efstathiou1992,Loeb2001,Benson2002,Okamoto2008,Gnedin2014}. These galaxies make up the population of ultrafaint dwarfs seen today as, for example, the smallest satellite galaxies orbiting the Milky Way \citep{Bose2018}. It should be noted, however, that dwarfs with masses similar to the Milky Way's ultrafaints, but which form in the {\it field}, may continue to form stars long after reionisation \citep{Bovill2011, Weisz2014}. In \galform{}, the effect of the global photoionising background is approximated by forbidding cooling in haloes of circular velocity, $V_c$, if $V_c < V_{{\rm cut}}$ at $z<z_{{\rm cut}}$. $V_{{\rm cut}}$ and $z_{{\rm cut}}$ are input parameters of the model and describe, respectively, the so-called ``filtering scale'' \citep{Gnedin2000,Benson2002} and the redshift of reionisation. Throughout, we assume $V_{{\rm cut}} = 30\,$kms$^{-1}$ and $z_{{\rm cut}} = 6$, as recommended by \cite{Bose2018}.

\begin{figure}
    \centering
    \includegraphics[width=\columnwidth]{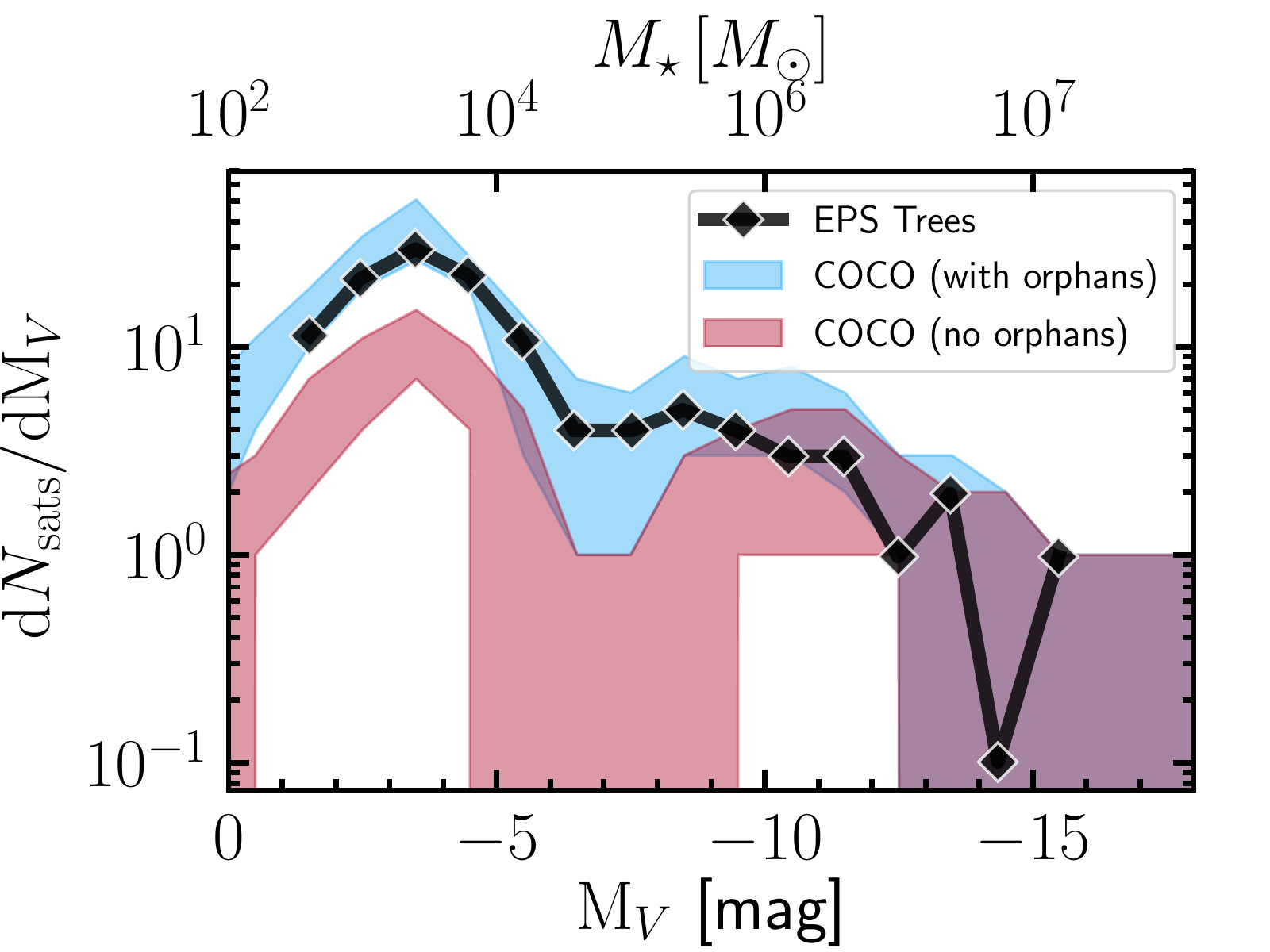}
    \caption{The average luminosity function of satellites in haloes of mass $M_{200}=1-1.3\times10^{12}\,{\rm M}_\odot$ measured in \coco{}, with (blue) and without (red) the inclusion of orphan galaxies. Shaded regions encompass the 10$^{{\rm th}}$ and 90$^{{\rm th}}$ percentiles. For comparison, the black line shows the mean luminosity function from 100 Monte Carlo merger tree realisations of haloes in the same mass range; the resolution of each merger tree is 100 times better than \coco{}. The inclusion of orphans significantly boosts the abundance of the smallest galaxies, starting at around $M_\star \sim10^6\,{\rm M}_\odot$. In particular, the calculation including orphans shows excellent agreement with the Monte Carlo trees, which demonstrates that tracking orphans mitigates the effects of limited numerical resolution in our $N$-body simulation.}
    \label{fig:orphan_conv}
\end{figure}

Finally, in \galform{} we also keep track of ``orphan galaxies'', objects whose dark matter haloes have been disrupted below \Subfind{}'s detection limit of 20 particles after falling into a more massive halo. In this situation, the galaxies in them are tracked using the methodology described in \cite{Simha2017}: in short, the subhaloes of these galaxies are followed up to the last output time at which they are resolved. Their future orbital evolution is captured by identifying these galaxies with the most-bound particle of the subhalo in which they originally formed. Thereafter, analytic formulae used to give the timescale for merging onto the central galaxy, accounting for the effects of dynamical friction and tidal disruption, are applied.

Fig.~\ref{fig:orphan_conv} shows the importance of keeping track of orphans when calculating the satellite population in haloes. Here we compare the average satellite luminosity function of Milky Way-mass haloes ($M_{200}=1-1.3\times10^{12}\,{\rm M}_\odot$; \citealt{Smith2007,Deason2012,Wang2015,Patel2018,Callingham2019,Deason2019,Grand2019}) with and without the inclusion of orphans. The former shows a significantly larger satellite population, particularly in the regime of the ultrafaint dwarfs ($M_\star\lesssim10^{5-6}\,{\rm M}_\odot$). The black curve shows the predictions from 100 Monte Carlo realisations of Galactic haloes generated using the extended Press-Schechter (EPS) formalism \citep{Press1974,Bond1991, Bower1991}: the minimum halo mass in each tree is roughly a factor of 100 smaller than in \coco{}\footnote{Note that the mass resolution in \coco{} is enough to resolve the atomic cooling limit ($V_c\sim17\,$kms$^{-1}$), so that any subhalo that can form stars through atomic cooling can do so. On the other hand, the parent \Color{} simulation just fails to resolve subhaloes near the atomic cooling limit. Thus, satellites in \coco{} make up a complete sample, whereas the larger volume \Color{} simulation provides enough statistical power to quantify qualitative trends identified in \coco{}.}.

The luminosity function including orphans is in excellent agreement with the much higher resolution calculation based on EPS trees. By contrast, a large number of ultrafaints are ``lost'' when counting only galaxies in resolved subhaloes at $z=0$ (see the curve without orphans). \cite{Newton2017} have shown that including orphans greatly improves convergence in the radial distribution of satellites between simulations of the same halo at different resolution; this is a central aspect of the results presented in Section~\ref{sect:disc}.

Tracking orphans also serves to compensate for any artificial disruption of subhaloes, that might be caused by numerical effects such as those discussed by \cite{vdB2018a} \& \cite{vdB2018b}, who suggest that subhaloes in cosmological simulations undergo excessive (unphysical) disruption, potentially underestimating the ``true'' subhalo population. Their observations are related to the findings from idealised, controlled simulations by \cite{Penarrubia2010} and \cite{Errani2019}, which suggest that ``cuspy'' dark matter subhaloes are, in general, resilient to the effects of tides. Since observed ultrafaints are unlikely to have undergone the type of violent baryonic effects that can produce cores \citep[e.g.][]{NEF96,Read2019}, their haloes are more likely to survive and remain cuspy. The inclusion of orphan galaxies mitigates, at least in part, the effects of artificial disruption. Finally, it is important to stress that the orphan galaxy scheme does not add `new' galaxies to the final catalogue, but simply keeps track of galaxies that had {\it already formed} in resolved subhaloes at earlier times and are therefore present in previous simulation output times.

\begin{figure*}
    \centering
    \includegraphics[width=\textwidth]{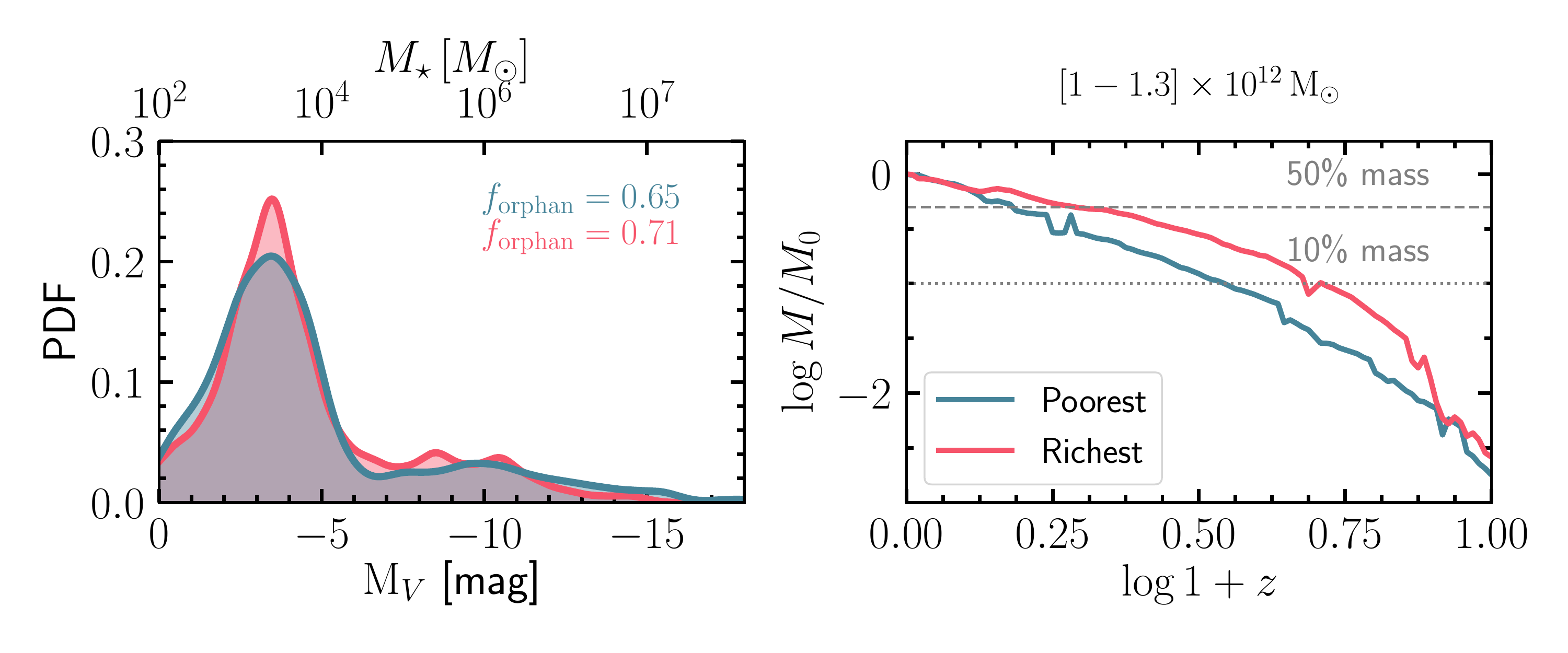}
    \caption{Correspondence between the mean satellite luminosity function (left panel) of Milky Way-mass haloes ($M_{200}^{z=0}\equiv M_0=1-1.3 \times 10^{12}\,{\rm M}_\odot$) and the mean assembly history of their host haloes (right panel). The figure displays these relations for the 20\% `richest' and 20\% `poorest' haloes in terms of the abundance of their ultrafaint populations. The right panel marks the epoch by which 10\% and 50\% of the present-day parent halo mass has collapsed. On average, early-forming Milky Way-analogues host more ultrafaint satellites than their later-forming counterparts. We have corrected the mean number of satellites for any differences in the mean final mass of host haloes in each of the `rich' and `poor' subcategories. Fig.~\ref{fig:orphan_compare_lf_assembly} compares this result with and without the inclusion of orphan galaxies.}
    \label{fig:mah_vs_uf}
\end{figure*}

\cite{Bose2018} provide a physical interpretation for the general form of the curves in Fig.~\ref{fig:orphan_conv} which they attribute to the way in which the process of reionisation shapes the present-day distribution of dwarf galaxies. In particular, the luminosity function of satellites consists of two sub-populations, an ultrafaint component and a bright component separated by a ``valley'' at ${\rm M}_V\approx-7$. The location of this kink is set by the filtering scale, corresponding to the choice of $V_{{\rm cut}} = 30\,$kms$^{-1}$ in the \galform{} model, which is itself motivated by analytic estimates \citep{Rees1986} and the results of hydrodynamical simulations \citep{Gnedin2000,Okamoto2008}. The ultrafaint galaxies, located to the left of this valley in Fig.~\ref{fig:orphan_conv}, assemble the bulk of their stellar mass prior to reionisation and are quenched thereafter; this is consistent with the star formation histories inferred from the ultrafaint satellites orbiting the Milky Way \citep{Brown2014}. Brighter galaxies form later, and their star formation histories bear little memory of reionisation (see also results from recent simulations by \citealt{Simpson2013}, \citealt{Wheeler2015} and \citealt{GarrisonKimmel2019}). Finally, the amplitude of the luminosity function faintwards of the valley depends sensitively on the redshift at which reionisation is complete. The bimodality of the satellite luminosity function in Fig.~\ref{fig:orphan_conv} is clearly present in the Milky Way data \citep{Bose2018}. In Section~\ref{sect:results}, we explore the connection between these satellite galaxies and the assembly histories of their host haloes. 

\section{Results}
\label{sect:results}

In the following subsections, we present the main results of our study. In Section~\ref{sect:assembly_number}, we demonstrate the dependence of the total satellite population on the formation time of host haloes, focussing, in Section~\ref{sect:milkyway}, on assembly histories similar to that of our own Galaxy. In Section~\ref{sect:assembly_radial} we study the influence of assembly history on the present-day radial distribution of satellite galaxies. Finally, in Section~\ref{sect:disc} we present a short discussion on corrections to these radial profiles after accounting for the destruction of satellite galaxies by the central galaxy.

\subsection{Dependence of the satellite population on the assembly histories of host haloes}
\label{sect:assembly_number}

We begin our investigation by considering the connection between the present-day dwarf galaxy content of Milky Way-mass haloes and their average mass accretion histories. In particular, we are interested in contrasting the differences, if any, between the mass assembly of haloes that are particularly ``rich'' or ``poor'' in their ultrafaint satellite content at $z=0$.

Predictions based on the \coco{} simulations are shown in Fig.~\ref{fig:mah_vs_uf}, which focuses on haloes in the mass range $M_{200}=\left[1-1.3\right]\times10^{12}\,{\rm M}_\odot$. Here, the distinction between the ``richest'' and ``poorest'' haloes, respectively, refers to the 20\% most abundant and the 20\% most deficient haloes in satellites of mass $M_\star \lesssim 10^5\,{\rm M}_\odot$ (see left panel). The mean accretion histories of the haloes, normalised to their present-day mass, are illustrated in the panels on the right. The offset between the two curves shows that, at fixed mass, haloes that form earlier tend to contain a larger population of ultrafaint satellites than their later-forming counterparts. The differences are larger at higher redshift: the ``richest'' and ``poorest'' haloes differ more in the epoch by which only 10\% of the final mass had been assembled than in the time when 50\% of the final mass was in place (a quantity often used to denote the redshift of formation of dark matter haloes). On average, early-forming haloes are $\sim30\%$ richer in their ultrafaint populations. On the other hand, the abundance of ``classical satellites'' (${\rm M}_V\geq-8.8$) is similar in the two classes of haloes.

The dependence of the number of ultrafaints on the formation time of haloes may be understood by considering the environmental dependence of halo merger rates at fixed mass. At fixed $z=0$ mass, early-forming haloes are likely ones originating from Lagrangian patches located in regions of high local overdensity \citep[e.g.][]{Sheth2004,Avila2005}. \cite{Fakhouri2009} have shown that at fixed mass, galactic haloes experience, on average, $\sim2.5\times$ more mergers (of all mass ratios) when located in the largest overdensities compared to their counterparts in underdensities. The increased number of mergers may subsequently translate into an increase in the abundance of satellites in early-forming haloes\footnote{Note, however, that the {\it total} mass accretion rate, which determines the final-day mass of the halo, is comparable for objects irrespective of their local environment: while accretion through discrete halo mergers is {\it positively} correlated with environment, the accretion rate of smooth, diffuse material is {\it negatively} correlated with halo environment \citep[][]{Fakhouri2010b}; the combined environmental dependence of the two modes of mass accretion therefore largely cancel out}.

\begin{figure*}
\centering
    \includegraphics[width=\textwidth]{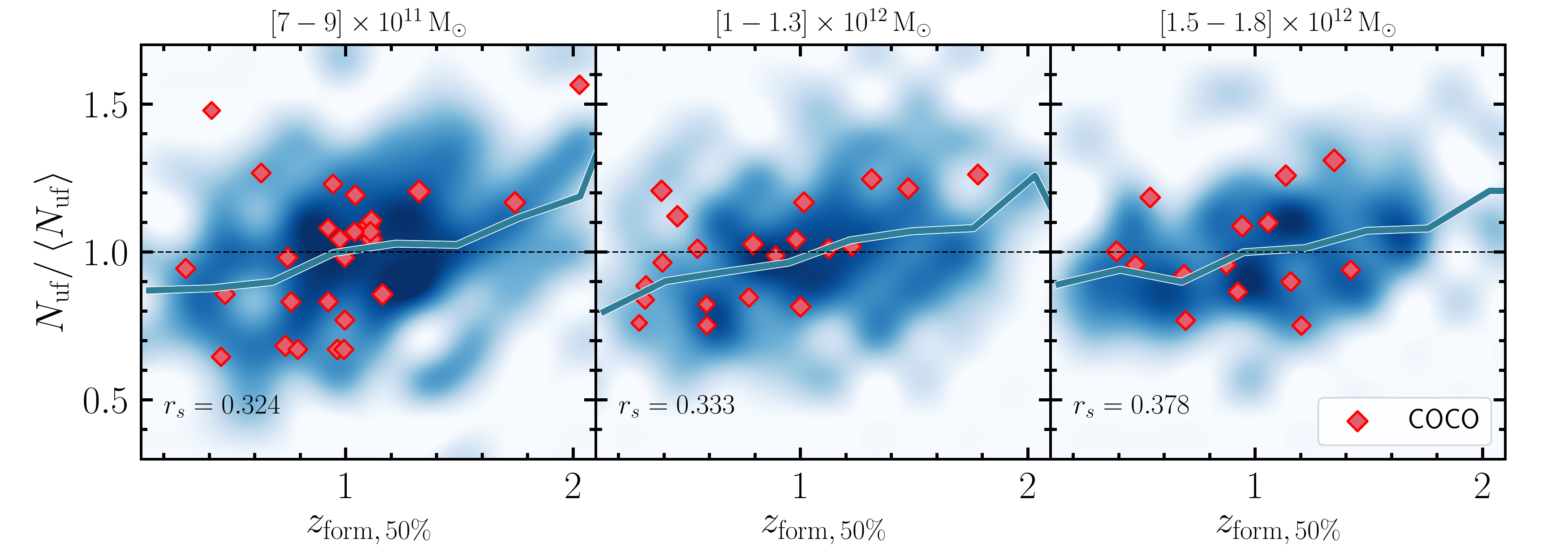}
    \caption{The number of ultrafaint dwarfs ($M_V \geq -6$; $M_\star \leq 10^5\,{\rm M}_\odot$) as a function of halo formation redshift, $z_{{\rm form}}$, defined as the time by which 50\% of the halo's final day mass was assembled. Each panel corresponds to a different bin of halo mass: $7-9 \times 10^{11}\,{\rm M}_\odot$ (left), $1-1.3 \times 10^{12}\,{\rm M}_\odot$ (middle) and $1.5-1.8 \times 10^{12}\,{\rm M}_\odot$ (right). Red diamonds represent results from \coco{}, with symbol sizes scaled by host halo mass. The blue cloud in the background of shows a 2D histogram of these quantities measured from \Color{}; the mean relation is indicated by the solid blue lines. In each panel, the number of ultrafaints is normalised to the mean number of ultrafaints hosted by haloes in that mass bin. We also give the Spearman rank correlation coefficient, $r_s$, in the bottom left corner of each panel.}
    \label{fig:Nuf_vs_z0p50}
\end{figure*}

\begin{figure*}
    \centering
    \includegraphics[width=\textwidth]{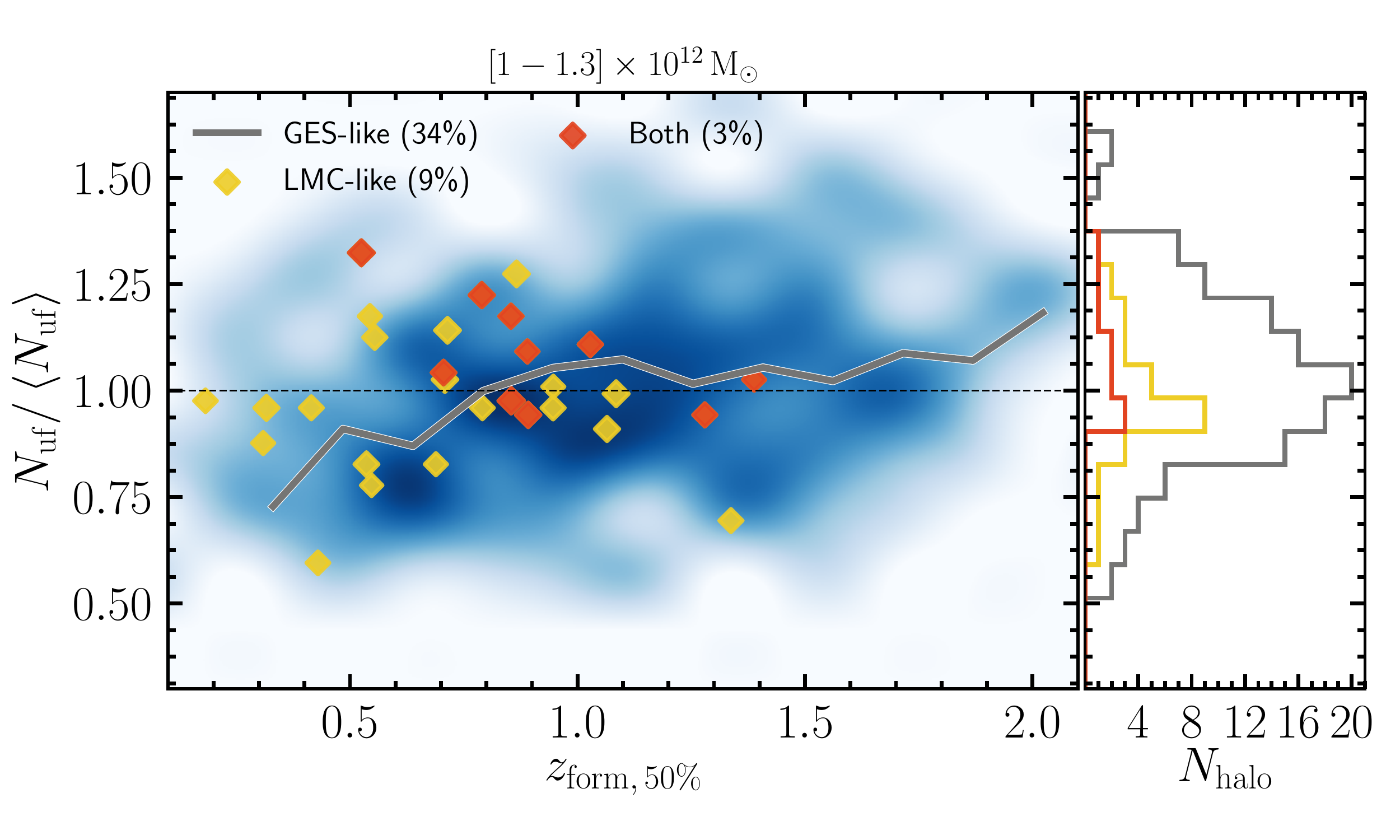}
    \caption{As Fig.~\ref{fig:Nuf_vs_z0p50}, but we now highlight sub-populations of haloes that have undergone accretion events characteristic of our own Galaxy, specifically, an early {\it Gaia}-Enceladus-Sausage-like (GES-like) accretion event (grey line, which shows the median trend with $z_{{\rm form},50\%}$), a late-time LMC-like accretion event (yellow) and both types of accretion event (red).}
    \label{fig:Nuf_vs_z0p50_mwspecific} 
\end{figure*}

Taken at face value, these results contradict the conclusions of
previous studies which have found that, in fact, early-forming haloes have {\it depleted} satellite populations \citep[e.g.][]{Gao2004,Zehavi2018,Artale2018,Bose2019}. However, there are two important distinctions between these studies and the present work: (1)~we are concerned with ultrafaint satellites whereas previous studies considered much brighter satellites ($M_\star\gtrsim10^7\,{\rm M}_\odot$), and (2)~the contribution of orphans was not taken into account in previous work. Indeed, relaxing requirements (1) and (2) in Fig.~\ref{fig:mah_vs_uf} reverts the correlation between formation   time and the abundance of satellites to the trends reported previously in the literature (see Appendix~\ref{sect:ImpOrphan}, Fig.~\ref{fig:orphan_compare_lf_assembly} for further details).

Fig.~\ref{fig:Nuf_vs_z0p50} shows the correlation between the number of ultrafaint satellites, $N_{{\rm uf}}$, and the formation epoch of the host halo more quantitatively. The diamonds represent individual haloes extracted from either \coco{} or \Color{}, and have been normalised to the mean number of ultrafaints, $\left<N_{{\rm uf}}\right>$, in the respective mass bin. Each panel corresponds to a different range of halo mass. Normalising $N_{{\rm uf}}$ to the mean value allows us to compare the predictions of \coco{} and \Color{}, despite the difference in mass resolution. This figure shows that there is indeed a positive correlation between the number of ultrafaints hosted by a halo and how early it formed, when the latter is characterised by the half-mass formation time, $z_{{\rm form,50\%}}$.

A simple way to quantify the strength of this correlation is through the Spearman rank correlation coefficient, $r_s$, which is quoted in the bottom left corner of each panel. An average score of $r_s\approx0.35$ suggests a weak, positive correlation between $N_{{\rm uf}}$ and $z_{{\rm form,50\%}}$. We find a similarly significant correlation between $N_{{\rm uf}}$ and $z_{{\rm form,10\%}}$, the redshift by which 10\% of the host halo's final mass was assembled (not shown here). On the other hand, $r_s\approx0$ for the correlation between $N_{{\rm uf}}$ and $z_{{\rm form,90\%}}$, the redshift by which 90\% of the host halo's final mass was in place (also not shown). This indicates that there is no significant correlation between the number of ultrafaints and the late-time accretion history of haloes.

In summary, these trends indicate that at fixed halo mass, the ultrafaint satellite content of Milky Way-mass haloes is correlated with the early accretion history of the halo, but is largely insensitive to its late-time merger history.

\subsection{The satellite content of haloes with Milky Way-like assembly histories}
\label{sect:milkyway}

In the previous subsection, we explored the connection between the richness of the ultrafaint satellite population of Milky Way-mass haloes and their accretion history, without specifying details of the events that determine the assembly of the halo. In this subsection, we consider the impact of the major accretion events thought to relevant for the assembly of our Galaxy.

Fig.~\ref{fig:Nuf_vs_z0p50_mwspecific} shows the dependence of $N_{{\rm uf}} / \left< N_{{\rm uf}} \right>$ on $z_{{\rm form,50\%}}$ for haloes in the mass range $M_{200}=\left[1-1.3\right]\times10^{12}\,{\rm M}_\odot$, where we have highlighted objects that have undergone distinct past accretion events in different colours. In particular, we show haloes with an LMC-like accretion event ($M_{{\rm halo}}^{{\rm infall}} \sim 10^{11}\,{\rm M}_\odot$ around 2 Gyrs ago, with the condition that at least one satellite as bright as the LMC is present at $z=0$, \citealt{Besla2015,Penarrubia2016}) in yellow, a {\it Gaia}-Enceladus-Sausage-like (GES-like) event ($M_{{\rm halo}}^{{\rm infall}} \sim 10^{11}\,{\rm M}_\odot$ between 8-10 Gyrs ago, \citealt{Belokurov2018,Helmi2018}) as a grey line (which shows the median relation trend with $z_{{\rm form},50\%}$); haloes that have experienced both types of accretion events are coloured in red. Note that we use symbols to represent individual haloes that fall in the LMC-like and LMC+GES-like categories as the statistics are not good enough to construct a meaningful median relation. In this figure, $M_{{\rm halo}}^{{\rm infall}}$ is the dark matter halo mass of the accreted satellite just prior to infall. The fraction of haloes that fall into each category is given at the top of this panel; the corresponding collapsed histograms of $N_{{\rm uf}}$ are shown in the right-hand panel. We present results from \Color{} only as the sample size in \coco{} is too small to split into the various categories. 

It is interesting to contrast the relative frequency of each type of merger. While an early GES-like accretion event is relatively common (roughly one in three haloes experience an event of this kind; see also \citealt{Fattahi2019}), a late-time LMC-like accretion event is less common (only around one in ten haloes experience this). The latter fraction is reduced by requiring that a satellite as bright as the LMC should survive to $z=0$ (the fraction increases to 28\% if this condition is relaxed). Assembly histories like that of our own Galaxy, where {\it both} types of events have occurred, are exceedingly rare: only 3\% of the haloes in this mass range in \Color{} fall in this category.

Fig.~\ref{fig:Nuf_vs_z0p50_mwspecific} also shows variations in the number of ultrafaint satellites in haloes that have experienced each type of accretion event. There is a roughly equal split of haloes above and below the mean population abundance in the category of LMC-like accretion events, while there is a marginal preference of an `ultrafaint excess' in haloes that have experienced an early GES-like accretion i.e., 56\% of haloes in this category lie above the one-to-one line, as evidenced by the extended tail $>1$ in the grey histogram. Interestingly, the distribution for haloes that have experienced both early and late-time accretion events (red diamonds) shows a tendency towards an excess of ultrafaint satellites relative to the mean population. This is especially true for lower values of $z_{{\rm form,50\%}}$.

The observations presented in Figs.~\ref{fig:mah_vs_uf},~\ref{fig:Nuf_vs_z0p50}~\&~\ref{fig:Nuf_vs_z0p50_mwspecific} paint a consistent picture: early-forming haloes contain, on average, a larger than average number of ultrafaint satellites, and an `ancient' GES-like accretion event is more influential in shaping the present-day satellite luminosity function than a late-time LMC-like accretion event.

\begin{figure*}
    \centering
    \includegraphics[width=0.475\textwidth]{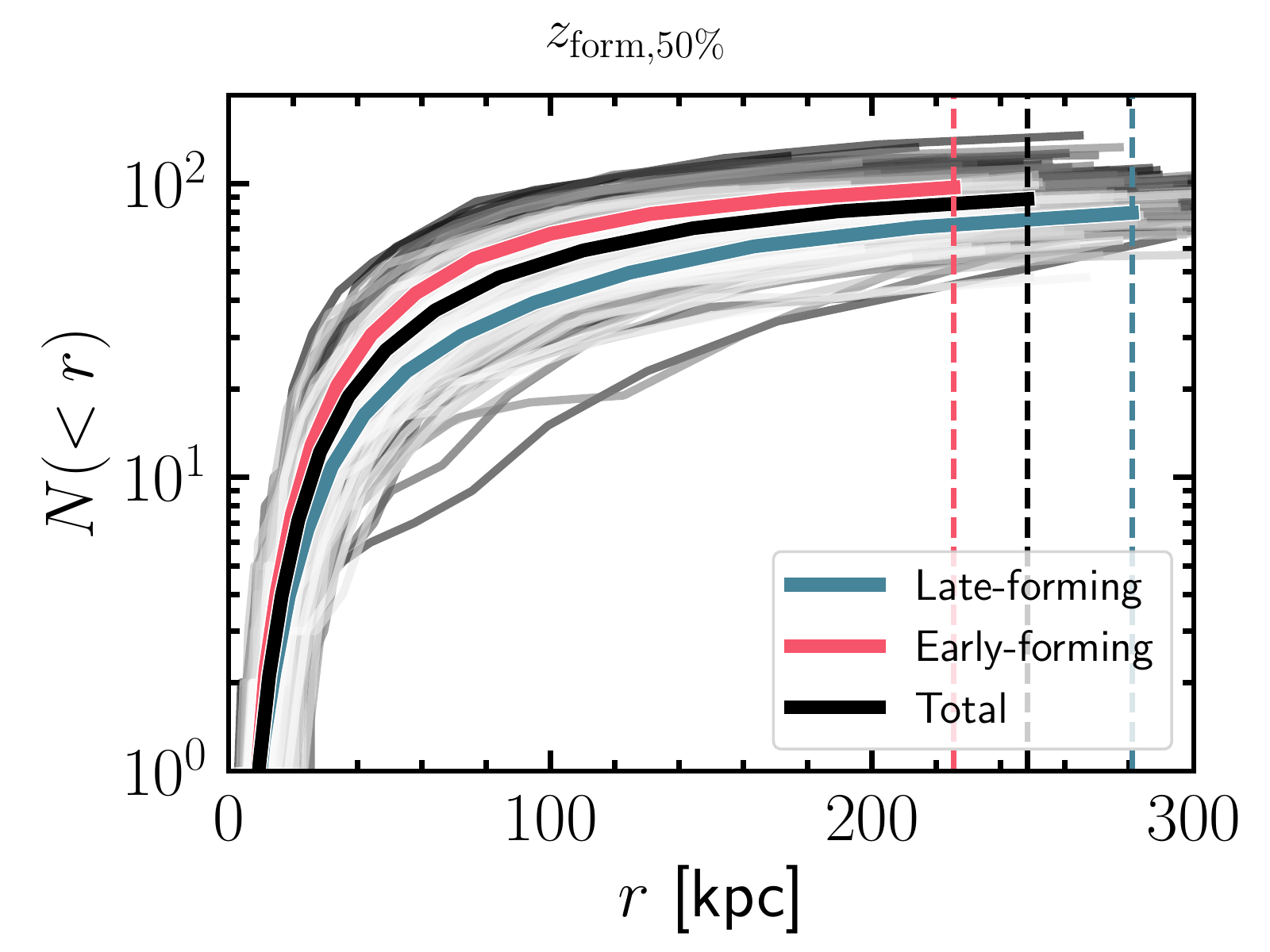}
    \includegraphics[width=0.475\textwidth]{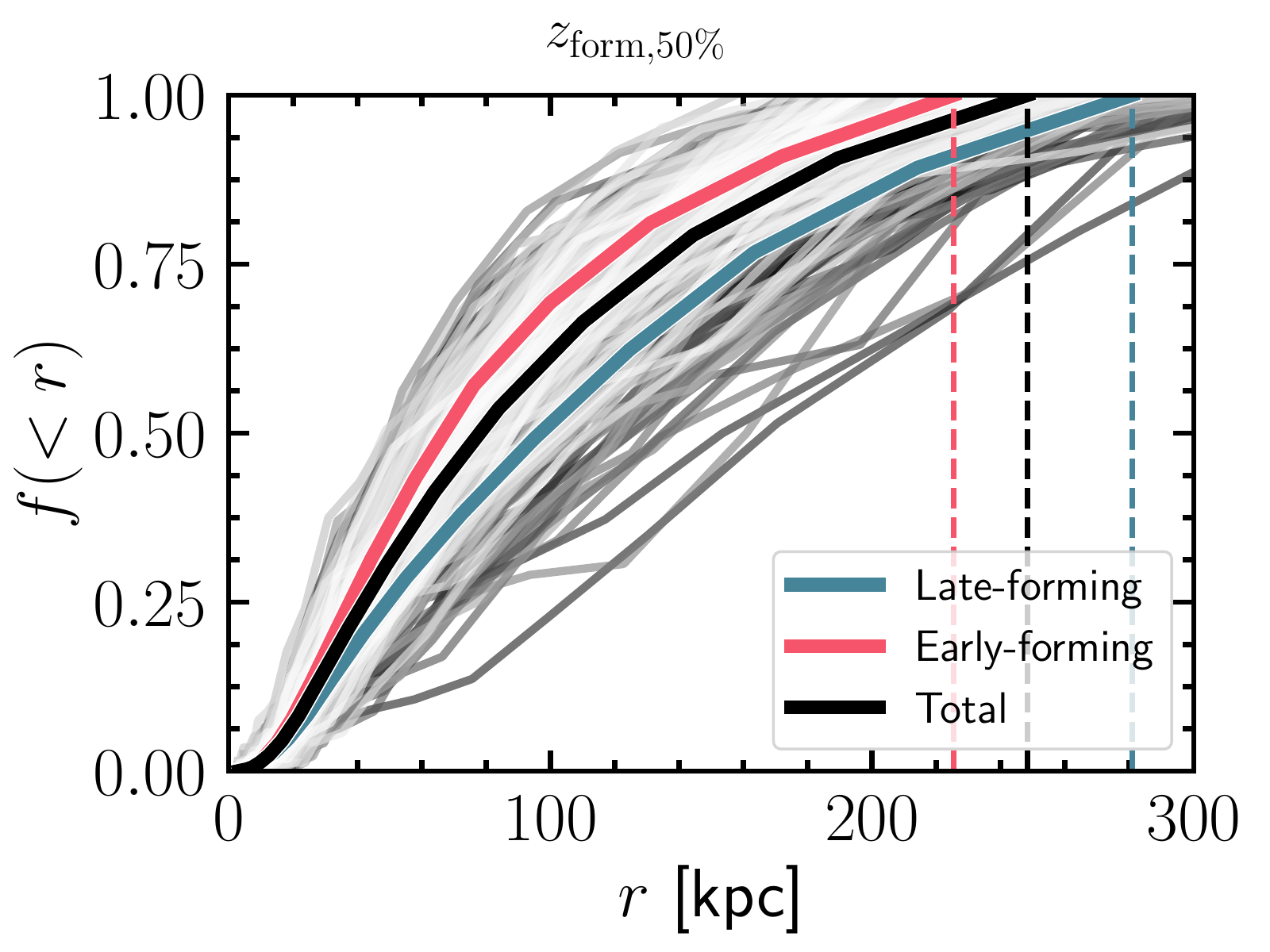}
        \caption{The radial distribution of satellites in galactic haloes ($M_{200} \sim 1-1.3\times10^{12}\,{\rm M}_\odot$) identified in the \Color{} simulation. {\bf Left panel}: the total number of satellites within distance, $r$, from the halo centre. {\bf Right panel}: the fraction of the total satellite population within this radius. In each panel, the grey curves show the radial profiles of individual haloes in this mass bin; the black curve is the corresponding mean profile. We distinguish between the profiles of the 20\% earliest- (red) and 20\% latest-forming (blue) haloes as defined by the redshift at which 50\% of the present-day halo mass was in place. The dashed vertical lines mark the corresponding virial radii. We note that the systematic differences observed in this figure are preserved when the radial profile is rescaled by the virial radius (i.e. when the $x$-axis is expressed as $r/r_{200}$) Neglecting orphan galaxies changes this radial distribution considerably; this comparison is presented in Fig.~\ref{fig:orphans_compare_radial_profile}.}
    \label{fig:radProf}
\end{figure*}

\subsection{Dependence of the radial distribution of satellites on the assembly histories of host haloes}
\label{sect:assembly_radial}

The radial distribution of satellites contains information about the formation and accretion history of the host halo and their dynamical evolution after infall. From a practical point of view, predictions from cosmological simulations for the radial distribution of satellites serve as an important prior when correcting for incompleteness in estimating the Milky Way's total satellite population from the partial observed set \citep[e.g.][]{Koposov2008,Tollerud2008,Belokurov2013,Hargis2014,Jethwa2016,Newton2017,Kim2018}.

The radial distribution of satellite galaxies after infall is the net result of a number of physical processes that interact non-trivially. Gravitational processes such as dynamical friction and tidal disruption are clearly important in determining the bias of the subhalo population relative to the underlying dark matter profile \citep[e.g.][]{Ghigna2000,Nagai2005,Diemand2007,Springel2008,Ludlow2009,Sawala2016,Han2016}, while the spatial bias of the galaxies relative to the smooth dark matter distribution and to subhaloes depends on the details of how galaxies occupy dark matter subhaloes \citep[e.g.][]{Frenk1996,Gao2004,vdB2005,Conroy2006,Maccio2010,Budzynski2012,Reddick2013}. In this subsection, we explore how the satellite distributions reflect differences in the assembly histories of haloes.

Fig.~\ref{fig:radProf} shows the radial profiles of satellites in hosts of mass $M_{200}=\left[1-1.3\right]\times10^{12}\,{\rm M}_\odot$ in \Color{}. The black curve shows the average radial profile, while the red line shows the average profile of the 20\% earliest-forming, and the blue line of the 20\% latest-forming, haloes in this mass bin, defined by their values of $z_{{\rm form,50\%}}$. The left panel shows the mean number of satellites identified within distance $r$ from the centre of the halo, while the right panel shows the fraction of the total satellite population located within this radius. The dashed vertical lines mark the mean virial radii of haloes in each category.

Fig.~\ref{fig:radProf} shows clear departures from the average profile when splitting haloes by their assembly time. Although early-forming haloes typically have lower masses (and smaller virial radii), we see that they tend to contain more satellites within a fixed physical halocentric radius than their later-forming counterparts. This is especially true within the innermost 100 kpc or so.  That late-forming haloes have a more spatially-extended satellite population is seen clearly in the right-hand panel of Fig.~\ref{fig:radProf}: while 75\% of the total satellite content of early-forming haloes is contained within 100 kpc of halo centre, only about 50\% of the late-forming population lies in this region. This is perhaps unsurprising: later-forming haloes are more likely to have undergone a more recent merger, in which case any satellites that have been brought in during this event are located preferentially in the outskirts of the host halo as dynamical friction has not operated for long enough to bring them to the centre. Simulations also show that satellites accreted later tend to have larger apocentres \citep[see Fig. 7 in][]{Deason2013}. This is likely the determining factor in the extended distribution of low-mass dwarf satellites for which dynamical friction is not very effective in any case. These qualitative trends change when orphan galaxies are not included, as shown in Fig.~\ref{fig:orphans_compare_radial_profile}.

\begin{figure}
    \centering
    \includegraphics[width=\columnwidth]{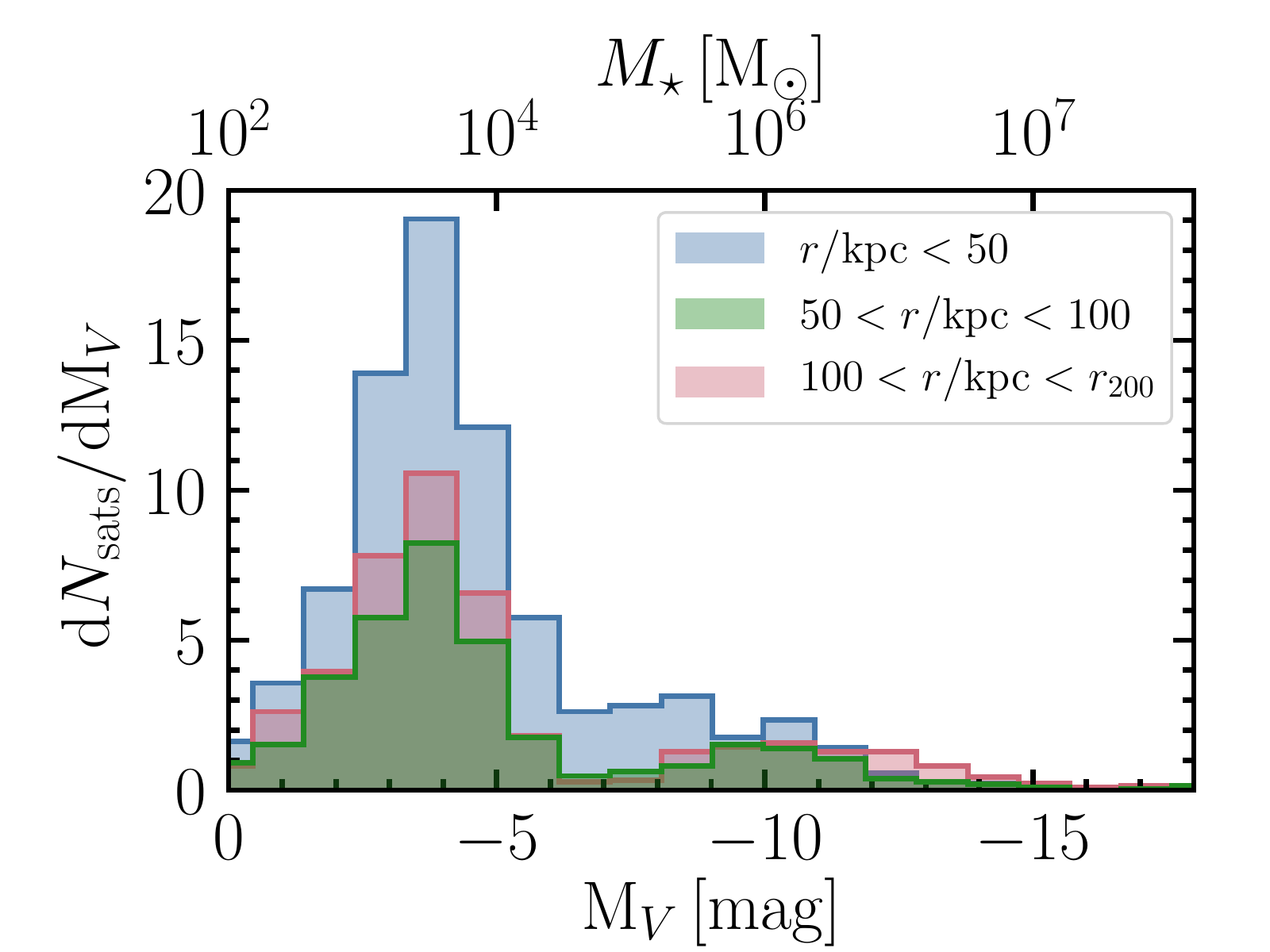}
    \caption{The luminosity function of satellites split into radial shells. The halo mass range adopted here corresponds to $M_{200} = 1-1.3\times10^{12}\,{\rm M}_\odot$. Whereas the classical satellites are found in roughly equal number throughout the halo, the ultrafaint population is found predominantly in the innermost 50 kpc. The effect of excluding the orphan galaxy population is shown in Fig.~\ref{fig:orphans_compare_lf_shells}.}
    \label{fig:lf_shells}
\end{figure}

Fig.~\ref{fig:lf_shells} splits the luminosity function of satellites into radial shells to highlight the mass range of satellites that dominates the population at a given radius. The classical dwarfs ($M_\star\geq10^6\,{\rm M}_\odot$) are distributed relatively evenly throughout the halo, although the brightest satellites are located preferentially in the outer parts. The ultrafaints, on the other hand, are more centrally-concentrated. As Fig.~\ref{fig:radProf} suggests, the ultrafaint population becomes more centrally-concentrated in early-forming host haloes. We note that it is in the inner regions of haloes where the effects of finite resolution are most important, and where the orphan galaxy tracking scheme is essential. In Fig.~\ref{fig:orphans_compare_lf_shells}, we show how that the radial occupation of satellites of a given mass changes significantly when orphans are not followed explicitly. 

\begin{figure*}
    \centering
    \includegraphics[width=\textwidth]{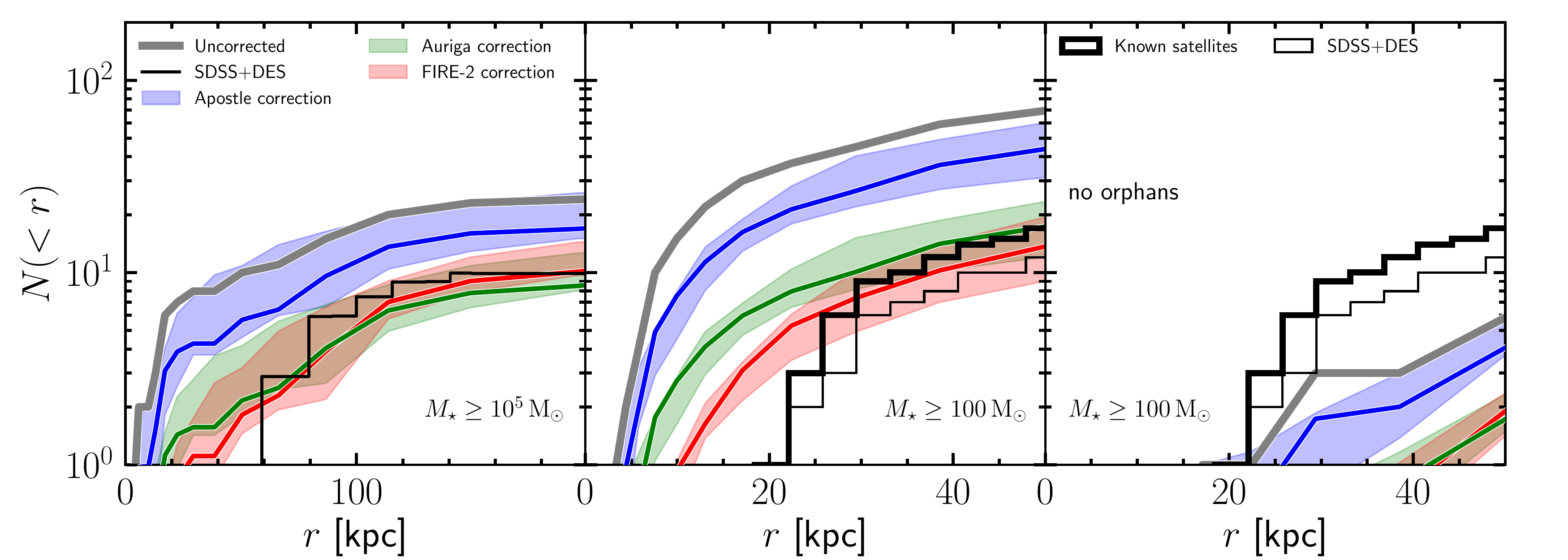}
    \caption{The radial distribution of satellites in galactic haloes from the \coco{} simulation. {\bf Left panel}: the mean radial distribution of satellites more massive than $M_\star\geq10^5\,{\rm M}_\odot$, before (grey line) and after (coloured bands) correcting for disc destruction. The size of the depletion correction varies from simulation to simulation: blue and green bands are the results found by \citet{Richings2018} in the {\sc Apostle} and {\sc Auriga} simulations respectively, and red bands are those found in {\sc FIRE-2} by \citet{Samuel2019}. The shaded regions encompass the 5$^{{\rm th}}$-95$^{{\rm th}}$ percentile scatter in the profiles measured in \coco{}. The black histogram shows the radial distribution of spectroscopically confirmed satellites in SDSS and DES that fall in this mass range. {\bf Middle panel}: a zoom into the radial distribution within the innermost 50 kpc, with linestyles and colours identical to those in the left panel. The figure now shows the radial distributions including all satellites more massive than $M_\star\geq100\,{\rm M}_\odot$. The thick black histogram is the distribution of {\em all known} satellites within 50 kpc, including those that have not been confirmed spectroscopically while the thin black histogram includes only those that have been confirmed. In both panels, the observational histograms have not been corrected for survey incompleteness, and should be treated as lower limits. In general, there is good agreement between the observed radial distribution and the simulated profiles after correcting for the effects of disc destruction. {\bf Right panel}: as in the middle panel, but neglecting `orphan' galaxies in the simulations. The exclusion of the orphan population reduces the satellite population dramatically; this is exacerbated when the effects of disc destruction are taken into account.}
    \label{fig:radDistr_zoombar}
\end{figure*}

\subsection{The destruction of satellites by the central disc}
\label{sect:disc}

In the previous subsections, we have described how the assembly histories of haloes give rise to differences in their present-day ultrafaint dwarf galaxy populations, both in number and in spatial distribution. We have especially highlighted the importance of tracking orphan galaxies when making these comparisons.

A well-known process that we now need to consider is the destruction of satellite galaxies resulting from interactions with the central baryonic disc. Both idealised and cosmological hydrodynamical simulations have shown that the amount of destruction is significant, particularly for satellites moving in radial orbits. The destruction is more severe near the centre and is enhanced around more massive central galaxies \citep[e.g.][]{Zentner2005,DOnghia2010,Yurin2015,Jethwa2016,Zhu2016,Sawala2017,Errani2017,GarrisonKimmel2017,Chua2017,Richings2018,Kelley2018,Simpson2018,Buck2019}.

\cite{Graus2018} have suggested that when disc disruption is accounted for, $\Lambda$CDM simulations are inconsistent with the abundance of satellites observed in the inner parts of the Milky Way. They find that under the assumption of ``standard reionisation quenching'' -- in which galaxies with peak circular velocity $V_{{\rm peak}}\leq20\,$kms$^{-1}$ no longer accrete gas after reionisation \citep{Okamoto2008,Okamoto2010} there are far fewer subhaloes within 50~kpc of the centre in the dark matter-only simulations than the observed number of satellites in the Milky Way. For their model to produce enough satellites in the inner galaxy, subhaloes as small as $V_{{\rm peak}}\simeq7\,$kms$^{-1}$ need to be populated with galaxies. As we shall see below the inclusion of orphan galaxies is enough to explain the Milky Way data within the conventional model of reionisation.

While the merger of satellites with the central baryonic disc is accounted for in \galform{}, the explicit destruction of satellites by interaction with the disc is not. We therefore correct the radial profiles in \galform{} using the fitting functions provided by \cite{Samuel2019}. These authors compare dark matter-only simulations with hydrodynamical simulations using the {\sc Fire-2} model \citep{Hopkins2018} and find that the relative difference, $f(d)$, in the corresponding radial profiles is well fit by the functional form:
\begin{equation}\label{eq:depletion}
f(d) = \begin{cases}
0\,, & 0\leq d \leq d_0\\
\alpha \left[ 1-\exp{\left(-\frac{d-d_0}{d_1}\right)} \right], & d\geq d_0  
\end{cases}
\end{equation}
where $d$ is the distance from the host galaxy and $\alpha$, $d_0$ and $d_1$ are free parameters that are fixed by fitting this function to the simulated radial profiles. This effect has also been quantified by \cite{Richings2018}, who performed a comprehensive analysis of the effects of disc destruction in the {\sc Apostle} \citep{Sawala2016,Fattahi2016} and {\sc Auriga} \citep{Grand2017} cosmological, hydrodynamical simulations. We find that the functional form in Eq.~\ref{eq:depletion} provides an equally good fit to the results of these simulations. The best-fitting values of $\alpha$, $d_0$ and $d_1$ for {\sc Fire-2}, {\sc Apostle} and {\sc Auriga} are listed in Table~\ref{tab:radial_fit_params}. 

The effects of disc destruction are illustrated in Fig.~\ref{fig:radDistr_zoombar}. The mean radial profile before applying any correction is shown by the solid grey line; the radial profiles after applying the corrections for disc destruction (using Eq.~\ref{eq:depletion}) are shown by the coloured bands, with each colour corresponding to the correction inferred from a different simulation: {\sc Fire-2} (red), {\sc Apostle} (blue) and {\sc Auriga} (green). Note that the radial distributions in this figure come from the high-resolution \coco{} simulation (rather than the lower resolution \Color{} simulation used for Fig.~\ref{fig:radProf}). The higher resolution of \coco{} is required to follow the very small ($M_\star\simeq100\,{\rm M}_\odot$) satellites shown in the right-hand panel of Fig.~\ref{fig:radDistr_zoombar}. We recall that the luminosity function of satellites in \coco{} gives a very good match to the ``resolution-free'' Monte Carlo calculation (see Fig.~\ref{fig:orphan_conv}).

Accounting for disc destruction clearly reduces the abundance of satellites in all cases. The reduction is particularly large within the innermost 80~kpc or so, especially for the {\sc Fire-2} model. The average number of satellites with $M_\star \geq 10^5\,{\rm M}_\odot$ within the virial radius is reduced, respectively, by 18\%, 56\% and 62\% when using the {\sc Apostle}, {\sc Auriga} and {\sc Fire-2}-based corrections. The differences amongst the different simulations arise primarily from the mass of the disc which is smaller in {\sc Apostle} than in {\sc Auriga} and {\sc Fire-2}. Further details of the reasons behind these differences are given in Appendix~\ref{sect:disc_disrupt_sims}. In general, the profiles predicted by \galform{} reproduce the rather flat distribution at
large radii characteristic of the Milky Way's radial satellite profile, extending out to $\sim300\,$kpc.

The middle panel in Fig.~\ref{fig:radDistr_zoombar} highlights the severity of disc disruption close to the galaxy by zooming into the innermost 50 kpc of the haloes. The black histograms show the radial distribution of observed satellites distinguishing the subset that have been spectroscopically confirmed as satellite galaxies in the SDSS and DES catalogues (thin black histogram) from the entire population of candidate satellites (thick black histogram). If disc destruction is neglected our model overpredicts the abundance of satellites within 50 kpc. However, once disc destruction is accounted for, the predicted radial profiles are in excellent agreement with the observations, at least when the corrections based on {\sc Auriga} and {\sc Fire-2} are applied. The histograms in the figure show that there are no observed satellites within $\sim 20$~kpc from the Galactic Centre; however, these data have not been corrected for sky coverage or survey incompleteness, and should therefore be treated as lower bounds on the `true' distribution. The {\sc Auriga} and {\sc Fire-2} models predict $\sim 7\pm 2$ and $4 \pm 2$ ultrafaint satellites within 20~kpc respectively.

The main conclusion drawn from Fig.~\ref{fig:radDistr_zoombar} is at odds with the results of \cite{Graus2018} who, after including the effects of disc destruction and assuming the standard model of reionisation, found about 10 times fewer satellites within 50 kpc of the halo centre than the observed number in this region of the Milky Way. The reason why \galform{} predicts far more ultrafaint satellites in the inner regions of galactic haloes is simply the inclusion of `orphan' galaxies, that is galaxies whose dark matter haloes have been lost due to numerical resolution effects as the number of particles in the subhalo drops below a certain level (20 particles in the case of \Subfind{} subhaloes). The resolution of the dark matter-only simulations used by \cite{Graus2018} is $3\times 10^4\, {\rm M}_\odot$. This is about a factor of 3 lower resolution than the `Level-2' Aquarius simulations of galactic haloes \citep{Springel2008} for which \cite{Newton2018} have shown orphan galaxies are important, particularly in the central regions. (They are important even at Aquarius `Level 1' resolution of $\sim 10^3 \, {\rm M}_\odot$.)

That neglecting orphan galaxies leads to a significant underestimate of the number of faint satellites can be seen in right-most panel of Fig.~\ref{fig:radDistr_zoombar}. The radial profiles that we find in this case are consistent with those of \cite{Graus2018}. Including the orphans obviates the need to populate extremely small subhaloes, as \cite{Graus2018} had to do in their simulations to obtain as many ultrafaint satellites as observed. In fact, in \galform{} the smallest haloes that ever form a galaxy have $V_{{\rm peak}} \sim 16.8\,$kms$^{-1}$. At $z=7$, when, according to \galform{}, these satellites have, on average, formed around 50\% of their $z=0$ stellar mass, this corresponds to a {\it halo mass} of $\sim1.4\times10^8\,{\rm M}_\odot$. A second important difference between our work and that of \cite{Graus2018} is the different assumed redshift of reionisation, $z=6$ in our model, $z=10$ in theirs. A later reionisation redshift boosts the abundance of ultrafaints (see Fig.~1 in \citealt{Bose2018} for a quantitative discussion). It is important to emphasise that the \galform{} model used here was not tuned to reproduce the observed radial distribution of satellites.

This discussion underlines the importance of limited numerical resolution in $N$-body simulations and the need to account for it, for example, using the orphan galaxy tracking scheme implemented in \galform{}. Including or ignoring these objects leads to very different inferences for the galaxy-halo connection in the dwarf galaxy regime.

\section{Conclusions}
\label{sect:conclusions}

We have used $N$-body simulations in the $\Lambda$CDM cosmology to explore the connection between the assembly history of Milky Way-mass dark matter haloes and their satellite population at $z=0$, including ``ultrafaint'' satellites and focusing on the abundance and spatial distribution of satellites.

Our sample of Milky Way-mass haloes was extracted from the dark matter-only {\it Copernicus Complexio Low Resolution} (\Color{}) and {\it Copernicus Complexio} (\coco{}) $N$-body simulations \citep{Sawala2016,Hellwing2016,Bose2016} of structure formation in a $\Lambda$CDM universe. The large computational volume ($10^6\,$Mpc$^3$) of the former provides a rich statistical sample of haloes of mass comparable to that of the Milky Way, whereas the 60 times higher mass resolution of \coco{} probes the smallest galaxies expected to form in haloes above the atomic gas cooling limit ($V_c\sim17\,$kms$^{-1}$). To embed galaxies within these simulations, we made use of the Durham semi-analytic model of galaxy formation, \galform{} \citep{Cole2000,Lacey2016}, which calculates the physical processes involved in galaxy formation along halo merger trees built from each simulation. \galform{} includes a detailed treatment of early hydrogen reionisation, gas cooling, star and black hole formation, feedback from stars and AGN, metal production, the synthesis of stellar populations, etc. It provides a flexible and computationally inexpensive environment to explore the parameter space of galaxy formation.

This work focuses on the demographics of the ultrafaint satellites of the Milky Way ($M_\star\lesssim10^5\,{\rm M}_\odot$, ${\rm M}_V\gtrsim-7$), and asks what their present-day abundance and radial distribution tells us about the assembly history of the host dark matter halo. These galaxies typically form in low-mass dark matter subhaloes, often near the detection threshold of the substructure finding algorithms that are applied to $N$-body simulations. To ensure that these galaxies are accounted for, even after their dark matter (sub)halo falls below the nominal resolution limit, we employ the technique of ``orphan galaxy tracking''. This method uses information from the last epoch at which the subhalo was resolved and follows its subsequent evolution by tagging its galaxy with the most-bound particle of the subhalo. Analytic prescriptions are then used to assess their survival under the influence of dynamical friction and tidal disruption \citep[e.g.][]{Simha2017}. The orphan method results in a dramatic improvement in the agreement between simulations at different resolution, as well as with Monte Carlo realisations of the galactic population which are not limited by numerical resolution (\citealt{Guo2011} and e.g. Fig.~1 of \citealt{Newton2017}). Our main results are summarised as follows:

\begin{enumerate}
    \item At fixed halo mass, that galactic dark matter haloes that form earlier than the average population of that mass contain more satellite galaxies than their later-forming counterparts (Fig.~\ref{fig:mah_vs_uf}).  
    \item The disparity in the total number of satellite galaxies between early and late-forming haloes is due primarily to ultrafaint satellites. The difference is manifest only when selecting haloes according to their early formation history (i.e., by the time when 10\% or 50\% of the present-day mass was in place). We find no correlation between the number of ultrafaint satellites and variations in the recent merger history of the host haloes (Fig.~\ref{fig:Nuf_vs_z0p50}).
    \item Translated into the language of the Milky Way's assembly, our models predict that an ancient {\it Gaia}-Enceladus-Sausage (GES)-like accretion event is more likely to have brought in a large number of ultrafaint galaxies than the more recent accretion of an LMC-like system. One plausible explanation for this is that an appreciable number of ultrafaint dwarfs may have already been destroyed inside the late-accreted host, diminishing its contribution to the Milky Way's satellite population. A full investigation would require tracking the fate of every satellite galaxy (and its progenitors) that ever falls into an LMC-mass halo; we leave this to future work. We also find that systems that have undergone {\it both} GES-like and LMC-like accretion events in their history are exceedingly rare: only $\sim3\%$ of the haloes in our sample have experienced GES and LMC-like mergers (Fig.~\ref{fig:Nuf_vs_z0p50_mwspecific}).
    \item The radial distribution of satellites is more centrally-concentrated in early-forming haloes. Haloes of a given present-day mass that assemble early have more centrally-concentrated matter density profiles, and the radial distribution of their satellite populations is similarly more concentrated. A (perhaps secondary) effect is that massive structures accreted early on (such as the GES progenitor), along with their satellites, are brought close to the central regions of the host by dynamical friction. The majority of these satellites are ultrafaint (Fig.~\ref{fig:radProf}). A sizeable fraction of these galaxies are identified as orphans (Fig.~\ref{fig:orphans_compare_radial_profile}).
    \item We accounted for the destruction of satellites by a central disc using the results of \cite{Samuel2019} for the {\sc Fire-2}, and those \cite{Richings2018} for the {\sc Apostle} and {\sc Auriga} hydrodynamical simulations (see  Appendix~\ref{sect:disc_disrupt_sims}). When orphans are not included, satellites are entirely missing within the inner 50 kpc of the halo, in severe tension with the observed radial distribution which has about 20 satellites in this region (right-hand panel of Fig.~\ref{fig:radDistr_zoombar}). Including orphans, however, brings the theoretical prediction into good agreement with the data  (middle panel of Fig.~\ref{fig:radDistr_zoombar}).

\end{enumerate}

In this work, we have explored the intimate connection between the present-day satellite population of galactic dark matter haloes and their formation history. The synergy of observational facilities such as the SDSS, DES and {\it Gaia} have transformed our understanding of the buildup of our own Galaxy. As we continue to extend the census of ultrafaint dwarfs around galaxies other than our own, the potential for developing a general picture of the assembly of galaxies becomes even greater. Theoretical models need to keep pace with the data; it is clear that the extreme ultrafaint dwarf galaxy regime continues to push the boundaries of state-of-the-art simulations, in resolution, quality of the galaxy formation physics models and subhalo tracking methods. Although this task is challenging, the wealth of information provided by these low-mass denizens, for both cosmology and galaxy formation, provides strong motivation to continue to strive for further advancements in observations, cosmological simulations and numerical methods.

\section*{Acknowledgements}
SB is supported by Harvard University through the ITC Fellowship. AJD is supported by a Royal Society University Research Fellowship, and she and CSF by the STFC Consolidated Grant for Astronomy at Durham (ST/L00075X/1). CSF is also supported by ERC Advanced Investigator grant, DMIDAS [GA 786910]. This work used the DiRAC Data Centric system at Durham University, operated by the Institute for Computational Cosmology on behalf of the STFC DiRAC HPC Facility (\href{www.dirac.ac.uk}{www.dirac.ac.uk}). This equipment was funded by BIS National E-infrastructure capital grant ST/K00042X/1, STFC capital grants ST/H008519/1 and ST/K00087X/1, STFC DiRAC Operations grant ST/K003267/1 and Durham University. DiRAC is part of the National E-Infrastructure. This research was carried out with the support of the HPC Infrastructure for Grand Challenges of Science and Engineering Project, co-financed by the European Regional Development Fund under the Innovative Economy Operational Programme. The work of SB was performed in part at Aspen Center for Physics, which is supported by National Science Foundation grant PHY-1607611. The data analysed in this paper can be made available upon request to the
author.

\bibliographystyle{mnras}
\bibliography{assembly}

\begin{thebibliography}{}
\makeatletter
\relax
\def\mn@urlcharsother{\let\do\@makeother \do\$\do\&\do\#\do\^\do\_\do\%\do\~}
\def\mn@doi{\begingroup\mn@urlcharsother \@ifnextchar [ {\mn@doi@}
  {\mn@doi@[]}}
\def\mn@doi@[#1]#2{\def\@tempa{#1}\ifx\@tempa\@empty \href
  {http://dx.doi.org/#2} {doi:#2}\else \href {http://dx.doi.org/#2} {#1}\fi
  \endgroup}
\def\mn@eprint#1#2{\mn@eprint@#1:#2::\@nil}
\def\mn@eprint@arXiv#1{\href {http://arxiv.org/abs/#1} {{\tt arXiv:#1}}}
\def\mn@eprint@dblp#1{\href {http://dblp.uni-trier.de/rec/bibtex/#1.xml}
  {dblp:#1}}
\def\mn@eprint@#1:#2:#3:#4\@nil{\def\@tempa {#1}\def\@tempb {#2}\def\@tempc
  {#3}\ifx \@tempc \@empty \let \@tempc \@tempb \let \@tempb \@tempa \fi \ifx
  \@tempb \@empty \def\@tempb {arXiv}\fi \@ifundefined
  {mn@eprint@\@tempb}{\@tempb:\@tempc}{\expandafter \expandafter \csname
  mn@eprint@\@tempb\endcsname \expandafter{\@tempc}}}

\bibitem[\protect\citeauthoryear{{Adelman-McCarthy} et~al.,}{{Adelman-McCarthy}
  et~al.}{2007}]{Adelman2007}
{Adelman-McCarthy} J.~K.,  et~al., 2007, \mn@doi [\apjs] {10.1086/518864},
  \href {http://adsabs.harvard.edu/abs/2007ApJS..172..634A} {172, 634}

\bibitem[\protect\citeauthoryear{{Alam} et~al.,}{{Alam}
  et~al.}{2015}]{Alam2015}
{Alam} S.,  et~al., 2015, \mn@doi [\apjs] {10.1088/0067-0049/219/1/12}, \href
  {http://adsabs.harvard.edu/abs/2015ApJS..219...12A} {219, 12}

\bibitem[\protect\citeauthoryear{{Artale}, {Zehavi}, {Contreras}  \&
  {Norberg}}{{Artale} et~al.}{2018}]{Artale2018}
{Artale} M.~C.,  {Zehavi} I.,  {Contreras} S.,   {Norberg} P.,  2018, \mn@doi
  [\mnras] {10.1093/mnras/sty2110}, \href
  {http://adsabs.harvard.edu/abs/2018MNRAS.480.3978A} {480, 3978}

\bibitem[\protect\citeauthoryear{{Avila-Reese}, {Col{\'{\i}}n},
  {Gottl{\"o}ber}, {Firmani}  \& {Maulbetsch}}{{Avila-Reese}
  et~al.}{2005}]{Avila2005}
{Avila-Reese} V.,  {Col{\'{\i}}n} P.,  {Gottl{\"o}ber} S.,  {Firmani} C.,
  {Maulbetsch} C.,  2005, \mn@doi [\apj] {10.1086/491726}, \href
  {http://adsabs.harvard.edu/abs/2005ApJ...634...51A} {634, 51}

\bibitem[\protect\citeauthoryear{{Baugh}, {Lacey}, {Frenk}, {Granato}, {Silva},
  {Bressan}, {Benson}  \& {Cole}}{{Baugh} et~al.}{2005}]{Baugh2005}
{Baugh} C.~M.,  {Lacey} C.~G.,  {Frenk} C.~S.,  {Granato} G.~L.,  {Silva} L.,
  {Bressan} A.,  {Benson} A.~J.,   {Cole} S.,  2005, \mn@doi [\mnras]
  {10.1111/j.1365-2966.2004.08553.x}, \href
  {http://adsabs.harvard.edu/abs/2005MNRAS.356.1191B} {356, 1191}

\bibitem[\protect\citeauthoryear{{Bechtol} et~al.,}{{Bechtol}
  et~al.}{2015}]{Bechtol2015}
{Bechtol} K.,  et~al., 2015, \mn@doi [\apj] {10.1088/0004-637X/807/1/50}, \href
  {http://adsabs.harvard.edu/abs/2015ApJ...807...50B} {807, 50}

\bibitem[\protect\citeauthoryear{{Belokurov}}{{Belokurov}}{2013}]{Belokurov2013}
{Belokurov} V.,  2013, \mn@doi [\nar] {10.1016/j.newar.2013.07.001}, \href
  {https://ui.adsabs.harvard.edu/abs/2013NewAR..57..100B} {57, 100}

\bibitem[\protect\citeauthoryear{{Belokurov}, {Erkal}, {Evans}, {Koposov}  \&
  {Deason}}{{Belokurov} et~al.}{2018}]{Belokurov2018}
{Belokurov} V.,  {Erkal} D.,  {Evans} N.~W.,  {Koposov} S.~E.,   {Deason}
  A.~J.,  2018, \mn@doi [\mnras] {10.1093/mnras/sty982}, \href
  {https://ui.adsabs.harvard.edu/abs/2018MNRAS.478..611B} {478, 611}

\bibitem[\protect\citeauthoryear{{Bennet}, {Sand}, {Crnojevi{\'c}}, {Spekkens},
  {Karunakaran}, {Zaritsky}  \& {Mutlu-Pakdil}}{{Bennet}
  et~al.}{2019}]{Bennet2019}
{Bennet} P.,  {Sand} D.~J.,  {Crnojevi{\'c}} D.,  {Spekkens} K.,  {Karunakaran}
  A.,  {Zaritsky} D.,   {Mutlu-Pakdil} B.,  2019, arXiv e-prints, \href
  {https://ui.adsabs.harvard.edu/abs/2019arXiv190603230B} {p. arXiv:1906.03230}

\bibitem[\protect\citeauthoryear{{Benson}}{{Benson}}{2012}]{Benson2012}
{Benson} A.~J.,  2012, \mn@doi [\na] {10.1016/j.newast.2011.07.004}, \href
  {http://adsabs.harvard.edu/abs/2012NewA...17..175B} {17, 175}

\bibitem[\protect\citeauthoryear{{Benson}, {Lacey}, {Baugh}, {Cole}  \&
  {Frenk}}{{Benson} et~al.}{2002a}]{Benson2002a}
{Benson} A.~J.,  {Lacey} C.~G.,  {Baugh} C.~M.,  {Cole} S.,   {Frenk} C.~S.,
  2002a, \mn@doi [\mnras] {10.1046/j.1365-8711.2002.05387.x}, \href
  {http://adsabs.harvard.edu/abs/2002MNRAS.333..156B} {333, 156}

\bibitem[\protect\citeauthoryear{{Benson}, {Frenk}, {Lacey}, {Baugh}  \&
  {Cole}}{{Benson} et~al.}{2002b}]{Benson2002}
{Benson} A.~J.,  {Frenk} C.~S.,  {Lacey} C.~G.,  {Baugh} C.~M.,   {Cole} S.,
  2002b, \mn@doi [\mnras] {10.1046/j.1365-8711.2002.05388.x}, \href
  {https://ui.adsabs.harvard.edu/abs/2002MNRAS.333..177B} {333, 177}

\bibitem[\protect\citeauthoryear{{Besla}}{{Besla}}{2015}]{Besla2015}
{Besla} G.,  2015, preprint, \href
  {http://adsabs.harvard.edu/abs/2015arXiv151103346B} {} (\mn@eprint {arXiv}
  {1511.03346})

\bibitem[\protect\citeauthoryear{{Besla}, {Kallivayalil}, {Hernquist},
  {Robertson}, {Cox}, {van der Marel}  \& {Alcock}}{{Besla}
  et~al.}{2007}]{Besla2007}
{Besla} G.,  {Kallivayalil} N.,  {Hernquist} L.,  {Robertson} B.,  {Cox} T.~J.,
   {van der Marel} R.~P.,   {Alcock} C.,  2007, \mn@doi [\apj]
  {10.1086/521385}, \href
  {https://ui.adsabs.harvard.edu/abs/2007ApJ...668..949B} {668, 949}

\bibitem[\protect\citeauthoryear{{Bond}, {Cole}, {Efstathiou}  \&
  {Kaiser}}{{Bond} et~al.}{1991}]{Bond1991}
{Bond} J.~R.,  {Cole} S.,  {Efstathiou} G.,   {Kaiser} N.,  1991, \mn@doi
  [\apj] {10.1086/170520}, \href
  {https://ui.adsabs.harvard.edu/abs/1991ApJ...379..440B} {379, 440}

\bibitem[\protect\citeauthoryear{{Bose}, {Hellwing}, {Frenk}, {Jenkins},
  {Lovell}, {Helly}  \& {Li}}{{Bose} et~al.}{2016}]{Bose2016}
{Bose} S.,  {Hellwing} W.~A.,  {Frenk} C.~S.,  {Jenkins} A.,  {Lovell} M.~R.,
  {Helly} J.~C.,   {Li} B.,  2016, \mn@doi [\mnras] {10.1093/mnras/stv2294},
  \href {http://adsabs.harvard.edu/abs/2016MNRAS.455..318B} {455, 318}

\bibitem[\protect\citeauthoryear{{Bose}, {Deason}  \& {Frenk}}{{Bose}
  et~al.}{2018}]{Bose2018}
{Bose} S.,  {Deason} A.~J.,   {Frenk} C.~S.,  2018, \mn@doi [\apj]
  {10.3847/1538-4357/aacbc4}, \href
  {https://ui.adsabs.harvard.edu/abs/2018ApJ...863..123B} {863, 123}

\bibitem[\protect\citeauthoryear{{Bose}, {Eisenstein}, {Hernquist},
  {Pillepich}, {Nelson}, {Marinacci}, {Springel}  \& {Vogelsberger}}{{Bose}
  et~al.}{2019}]{Bose2019}
{Bose} S.,  {Eisenstein} D.~J.,  {Hernquist} L.,  {Pillepich} A.,  {Nelson} D.,
   {Marinacci} F.,  {Springel} V.,   {Vogelsberger} M.,  2019, arXiv e-prints,
  \href {https://ui.adsabs.harvard.edu/abs/2019arXiv190508799B} {p.
  arXiv:1905.08799}

\bibitem[\protect\citeauthoryear{{Bovill} \& {Ricotti}}{{Bovill} \&
  {Ricotti}}{2009}]{Bovill2009}
{Bovill} M.~S.,  {Ricotti} M.,  2009, \mn@doi [\apj]
  {10.1088/0004-637X/693/2/1859}, \href
  {https://ui.adsabs.harvard.edu/abs/2009ApJ...693.1859B} {693, 1859}

\bibitem[\protect\citeauthoryear{{Bovill} \& {Ricotti}}{{Bovill} \&
  {Ricotti}}{2011}]{Bovill2011}
{Bovill} M.~S.,  {Ricotti} M.,  2011, \mn@doi [\apj]
  {10.1088/0004-637X/741/1/18}, \href
  {https://ui.adsabs.harvard.edu/abs/2011ApJ...741...18B} {741, 18}

\bibitem[\protect\citeauthoryear{{Bower}}{{Bower}}{1991}]{Bower1991}
{Bower} R.~G.,  1991, \mn@doi [\mnras] {10.1093/mnras/248.2.332}, \href
  {https://ui.adsabs.harvard.edu/abs/1991MNRAS.248..332B} {248, 332}

\bibitem[\protect\citeauthoryear{{Bower}, {Benson}, {Malbon}, {Helly}, {Frenk},
  {Baugh}, {Cole}  \& {Lacey}}{{Bower} et~al.}{2006}]{Bower2006}
{Bower} R.~G.,  {Benson} A.~J.,  {Malbon} R.,  {Helly} J.~C.,  {Frenk} C.~S.,
  {Baugh} C.~M.,  {Cole} S.,   {Lacey} C.~G.,  2006, \mn@doi [\mnras]
  {10.1111/j.1365-2966.2006.10519.x}, \href
  {http://adsabs.harvard.edu/abs/2006MNRAS.370..645B} {370, 645}

\bibitem[\protect\citeauthoryear{{Boylan-Kolchin}, {Besla}  \&
  {Hernquist}}{{Boylan-Kolchin} et~al.}{2011}]{BoylanKolchin2011}
{Boylan-Kolchin} M.,  {Besla} G.,   {Hernquist} L.,  2011, \mn@doi [\mnras]
  {10.1111/j.1365-2966.2011.18495.x}, \href
  {https://ui.adsabs.harvard.edu/abs/2011MNRAS.414.1560B} {414, 1560}

\bibitem[\protect\citeauthoryear{{Boylan-Kolchin}, {Bullock}, {Sohn}, {Besla}
  \& {van der Marel}}{{Boylan-Kolchin} et~al.}{2013}]{BoylanKolchin2013}
{Boylan-Kolchin} M.,  {Bullock} J.~S.,  {Sohn} S.~T.,  {Besla} G.,   {van der
  Marel} R.~P.,  2013, \mn@doi [\apj] {10.1088/0004-637X/768/2/140}, \href
  {http://adsabs.harvard.edu/abs/2013ApJ...768..140B} {768, 140}

\bibitem[\protect\citeauthoryear{{Brooks} \& {Zolotov}}{{Brooks} \&
  {Zolotov}}{2014}]{Brooks2014}
{Brooks} A.~M.,  {Zolotov} A.,  2014, \mn@doi [\apj]
  {10.1088/0004-637X/786/2/87}, \href
  {https://ui.adsabs.harvard.edu/abs/2014ApJ...786...87B} {786, 87}

\bibitem[\protect\citeauthoryear{{Brown} et~al.,}{{Brown}
  et~al.}{2014}]{Brown2014}
{Brown} T.~M.,  et~al., 2014, \mn@doi [\apj] {10.1088/0004-637X/796/2/91},
  \href {https://ui.adsabs.harvard.edu/abs/2014ApJ...796...91B} {796, 91}

\bibitem[\protect\citeauthoryear{{Buck}, {Macci{\`o}}, {Dutton}, {Obreja}  \&
  {Frings}}{{Buck} et~al.}{2019}]{Buck2019}
{Buck} T.,  {Macci{\`o}} A.~V.,  {Dutton} A.~A.,  {Obreja} A.,   {Frings} J.,
  2019, \mn@doi [\mnras] {10.1093/mnras/sty2913}, \href
  {https://ui.adsabs.harvard.edu/abs/2019MNRAS.483.1314B} {483, 1314}

\bibitem[\protect\citeauthoryear{{Budzynski}, {Koposov}, {McCarthy}, {McGee}
  \& {Belokurov}}{{Budzynski} et~al.}{2012}]{Budzynski2012}
{Budzynski} J.~M.,  {Koposov} S.~E.,  {McCarthy} I.~G.,  {McGee} S.~L.,
  {Belokurov} V.,  2012, \mn@doi [\mnras] {10.1111/j.1365-2966.2012.20663.x},
  \href {https://ui.adsabs.harvard.edu/abs/2012MNRAS.423..104B} {423, 104}

\bibitem[\protect\citeauthoryear{{Bullock}, {Kravtsov}  \&
  {Weinberg}}{{Bullock} et~al.}{2000}]{Bullock2000}
{Bullock} J.~S.,  {Kravtsov} A.~V.,   {Weinberg} D.~H.,  2000, \mn@doi [\apj]
  {10.1086/309279}, \href {http://adsabs.harvard.edu/abs/2000ApJ...539..517B}
  {539, 517}

\bibitem[\protect\citeauthoryear{{Callingham} et~al.,}{{Callingham}
  et~al.}{2019}]{Callingham2019}
{Callingham} T.~M.,  et~al., 2019, \mn@doi [\mnras] {10.1093/mnras/stz365},
  \href {https://ui.adsabs.harvard.edu/abs/2019MNRAS.484.5453C} {484, 5453}

\bibitem[\protect\citeauthoryear{{Chambers} et~al.,}{{Chambers}
  et~al.}{2016}]{Chambers2016}
{Chambers} K.~C.,  et~al., 2016, preprint, \href
  {http://adsabs.harvard.edu/abs/2016arXiv161205560C} {} (\mn@eprint {arXiv}
  {1612.05560})

\bibitem[\protect\citeauthoryear{{Chua}, {Pillepich}, {Rodriguez-Gomez},
  {Vogelsberger}, {Bird}  \& {Hernquist}}{{Chua} et~al.}{2017}]{Chua2017}
{Chua} K.~T.~E.,  {Pillepich} A.,  {Rodriguez-Gomez} V.,  {Vogelsberger} M.,
  {Bird} S.,   {Hernquist} L.,  2017, \mn@doi [\mnras] {10.1093/mnras/stx2238},
  \href {http://adsabs.harvard.edu/abs/2017MNRAS.472.4343C} {472, 4343}

\bibitem[\protect\citeauthoryear{{Cole}, {Aragon-Salamanca}, {Frenk}, {Navarro}
   \& {Zepf}}{{Cole} et~al.}{1994}]{Cole1994}
{Cole} S.,  {Aragon-Salamanca} A.,  {Frenk} C.~S.,  {Navarro} J.~F.,   {Zepf}
  S.~E.,  1994, \mn@doi [\mnras] {10.1093/mnras/271.4.781}, \href
  {http://adsabs.harvard.edu/abs/1994MNRAS.271..781C} {271, 781}

\bibitem[\protect\citeauthoryear{{Cole}, {Lacey}, {Baugh}  \& {Frenk}}{{Cole}
  et~al.}{2000}]{Cole2000}
{Cole} S.,  {Lacey} C.~G.,  {Baugh} C.~M.,   {Frenk} C.~S.,  2000, \mn@doi
  [\mnras] {10.1046/j.1365-8711.2000.03879.x}, \href
  {http://adsabs.harvard.edu/abs/2000MNRAS.319..168C} {319, 168}

\bibitem[\protect\citeauthoryear{{Conroy}, {Wechsler}  \& {Kravtsov}}{{Conroy}
  et~al.}{2006}]{Conroy2006}
{Conroy} C.,  {Wechsler} R.~H.,   {Kravtsov} A.~V.,  2006, \mn@doi [\apj]
  {10.1086/503602}, \href {http://adsabs.harvard.edu/abs/2006ApJ...647..201C}
  {647, 201}

\bibitem[\protect\citeauthoryear{{Correa}, {Wyithe}, {Schaye}  \&
  {Duffy}}{{Correa} et~al.}{2015}]{Correa2015}
{Correa} C.~A.,  {Wyithe} J. S.~B.,  {Schaye} J.,   {Duffy} A.~R.,  2015,
  \mn@doi [\mnras] {10.1093/mnras/stv689}, \href
  {https://ui.adsabs.harvard.edu/abs/2015MNRAS.450.1514C} {450, 1514}

\bibitem[\protect\citeauthoryear{{Couchman} \& {Rees}}{{Couchman} \&
  {Rees}}{1986}]{Couchman1986}
{Couchman} H.~M.~P.,  {Rees} M.~J.,  1986, \mn@doi [\mnras]
  {10.1093/mnras/221.1.53}, \href
  {http://adsabs.harvard.edu/abs/1986MNRAS.221...53C} {221, 53}

\bibitem[\protect\citeauthoryear{{Croton} et~al.,}{{Croton}
  et~al.}{2006}]{Croton2006}
{Croton} D.~J.,  et~al., 2006, \mn@doi [\mnras]
  {10.1111/j.1365-2966.2005.09675.x}, \href
  {http://adsabs.harvard.edu/abs/2006MNRAS.365...11C} {365, 11}

\bibitem[\protect\citeauthoryear{{D'Onghia}, {Springel}, {Hernquist}  \&
  {Keres}}{{D'Onghia} et~al.}{2010}]{DOnghia2010}
{D'Onghia} E.,  {Springel} V.,  {Hernquist} L.,   {Keres} D.,  2010, \mn@doi
  [\apj] {10.1088/0004-637X/709/2/1138}, \href
  {http://adsabs.harvard.edu/abs/2010ApJ...709.1138D} {709, 1138}

\bibitem[\protect\citeauthoryear{{Davis}, {Efstathiou}, {Frenk}  \&
  {White}}{{Davis} et~al.}{1985}]{Davis1985}
{Davis} M.,  {Efstathiou} G.,  {Frenk} C.~S.,   {White} S.~D.~M.,  1985,
  \mn@doi [\apj] {10.1086/163168}, \href
  {http://adsabs.harvard.edu/abs/1985ApJ...292..371D} {292, 371}

\bibitem[\protect\citeauthoryear{{Deason} et~al.,}{{Deason}
  et~al.}{2012}]{Deason2012}
{Deason} A.~J.,  et~al., 2012, \mn@doi [\mnras]
  {10.1111/j.1365-2966.2012.21639.x}, \href
  {http://adsabs.harvard.edu/abs/2012MNRAS.425.2840D} {425, 2840}

\bibitem[\protect\citeauthoryear{{Deason}, {Belokurov}, {Evans}  \&
  {Johnston}}{{Deason} et~al.}{2013}]{Deason2013}
{Deason} A.~J.,  {Belokurov} V.,  {Evans} N.~W.,   {Johnston} K.~V.,  2013,
  \mn@doi [\apj] {10.1088/0004-637X/763/2/113}, \href
  {https://ui.adsabs.harvard.edu/abs/2013ApJ...763..113D} {763, 113}

\bibitem[\protect\citeauthoryear{{Deason}, {Wetzel}, {Garrison-Kimmel}  \&
  {Belokurov}}{{Deason} et~al.}{2015}]{Deason2015}
{Deason} A.~J.,  {Wetzel} A.~R.,  {Garrison-Kimmel} S.,   {Belokurov} V.,
  2015, \mn@doi [\mnras] {10.1093/mnras/stv1939}, \href
  {https://ui.adsabs.harvard.edu/abs/2015MNRAS.453.3568D} {453, 3568}

\bibitem[\protect\citeauthoryear{{Deason}, {Fattahi}, {Belokurov}, {Evans},
  {Grand}, {Marinacci}  \& {Pakmor}}{{Deason} et~al.}{2019}]{Deason2019}
{Deason} A.~J.,  {Fattahi} A.,  {Belokurov} V.,  {Evans} N.~W.,  {Grand} R.
  J.~J.,  {Marinacci} F.,   {Pakmor} R.,  2019, \mn@doi [\mnras]
  {10.1093/mnras/stz623}, \href
  {https://ui.adsabs.harvard.edu/abs/2019MNRAS.485.3514D} {485, 3514}

\bibitem[\protect\citeauthoryear{{Diemand}, {Kuhlen}  \& {Madau}}{{Diemand}
  et~al.}{2007}]{Diemand2007}
{Diemand} J.,  {Kuhlen} M.,   {Madau} P.,  2007, \mn@doi [\apj]
  {10.1086/520573}, \href
  {https://ui.adsabs.harvard.edu/abs/2007ApJ...667..859D} {667, 859}

\bibitem[\protect\citeauthoryear{{Dooley}, {Peter}, {Carlin}, {Frebel},
  {Bechtol}  \& {Willman}}{{Dooley} et~al.}{2017}]{Dooley2017}
{Dooley} G.~A.,  {Peter} A.~H.~G.,  {Carlin} J.~L.,  {Frebel} A.,  {Bechtol}
  K.,   {Willman} B.,  2017, \mn@doi [\mnras] {10.1093/mnras/stx2001}, \href
  {http://adsabs.harvard.edu/abs/2017MNRAS.472.1060D} {472, 1060}

\bibitem[\protect\citeauthoryear{{Doroshkevich}, {Zel'dovich}  \&
  {Novikov}}{{Doroshkevich} et~al.}{1967}]{Doroshkevich1967}
{Doroshkevich} A.~G.,  {Zel'dovich} Y.~B.,   {Novikov} I.~D.,  1967, \sovast,
  \href {http://adsabs.harvard.edu/abs/1967SvA....11..233D} {11, 233}

\bibitem[\protect\citeauthoryear{{Drlica-Wagner} et~al.,}{{Drlica-Wagner}
  et~al.}{2015}]{Drlica2015}
{Drlica-Wagner} A.,  et~al., 2015, \mn@doi [\apj]
  {10.1088/0004-637X/813/2/109}, \href
  {http://adsabs.harvard.edu/abs/2015ApJ...813..109D} {813, 109}

\bibitem[\protect\citeauthoryear{{Efstathiou}}{{Efstathiou}}{1992}]{Efstathiou1992}
{Efstathiou} G.,  1992, \mn@doi [\mnras] {10.1093/mnras/256.1.43P}, \href
  {http://adsabs.harvard.edu/abs/1992MNRAS.256P..43E} {256, 43P}

\bibitem[\protect\citeauthoryear{{Errani} \& {Pe{\~n}arrubia}}{{Errani} \&
  {Pe{\~n}arrubia}}{2019}]{Errani2019}
{Errani} R.,  {Pe{\~n}arrubia} J.,  2019, arXiv e-prints, \href
  {https://ui.adsabs.harvard.edu/abs/2019arXiv190601642E} {p. arXiv:1906.01642}

\bibitem[\protect\citeauthoryear{{Errani}, {Pe{\~n}arrubia}, {Laporte}  \&
  {G{\'o}mez}}{{Errani} et~al.}{2017}]{Errani2017}
{Errani} R.,  {Pe{\~n}arrubia} J.,  {Laporte} C. F.~P.,   {G{\'o}mez} F.~A.,
  2017, \mn@doi [Monthly Notices of the Royal Astronomical Society]
  {10.1093/mnrasl/slw211}, \href
  {https://ui.adsabs.harvard.edu/abs/2017MNRAS.465L..59E} {465, L59}

\bibitem[\protect\citeauthoryear{{Fakhouri} \& {Ma}}{{Fakhouri} \&
  {Ma}}{2009}]{Fakhouri2009}
{Fakhouri} O.,  {Ma} C.-P.,  2009, \mn@doi [\mnras]
  {10.1111/j.1365-2966.2009.14480.x}, \href
  {https://ui.adsabs.harvard.edu/abs/2009MNRAS.394.1825F} {394, 1825}

\bibitem[\protect\citeauthoryear{{Fakhouri} \& {Ma}}{{Fakhouri} \&
  {Ma}}{2010}]{Fakhouri2010b}
{Fakhouri} O.,  {Ma} C.-P.,  2010, \mn@doi [\mnras]
  {10.1111/j.1365-2966.2009.15844.x}, \href
  {https://ui.adsabs.harvard.edu/abs/2010MNRAS.401.2245F} {401, 2245}

\bibitem[\protect\citeauthoryear{{Fakhouri}, {Ma}  \&
  {Boylan-Kolchin}}{{Fakhouri} et~al.}{2010}]{Fakhouri2010}
{Fakhouri} O.,  {Ma} C.-P.,   {Boylan-Kolchin} M.,  2010, \mn@doi [\mnras]
  {10.1111/j.1365-2966.2010.16859.x}, \href
  {https://ui.adsabs.harvard.edu/abs/2010MNRAS.406.2267F} {406, 2267}

\bibitem[\protect\citeauthoryear{{Fattahi} et~al.,}{{Fattahi}
  et~al.}{2016}]{Fattahi2016}
{Fattahi} A.,  et~al., 2016, \mn@doi [\mnras] {10.1093/mnras/stv2970}, \href
  {http://adsabs.harvard.edu/abs/2016MNRAS.457..844F} {457, 844}

\bibitem[\protect\citeauthoryear{{Fattahi} et~al.,}{{Fattahi}
  et~al.}{2019}]{Fattahi2019}
{Fattahi} A.,  et~al., 2019, \mn@doi [\mnras] {10.1093/mnras/stz159}, \href
  {https://ui.adsabs.harvard.edu/abs/2019MNRAS.484.4471F} {484, 4471}

\bibitem[\protect\citeauthoryear{{Font} et~al.,}{{Font}
  et~al.}{2011}]{Font2011}
{Font} A.~S.,  et~al., 2011, \mn@doi [\mnras]
  {10.1111/j.1365-2966.2011.19339.x}, \href
  {https://ui.adsabs.harvard.edu/abs/2011MNRAS.417.1260F} {417, 1260}

\bibitem[\protect\citeauthoryear{{Frenk}, {White}, {Efstathiou}  \&
  {Davis}}{{Frenk} et~al.}{1985}]{Frenk1985}
{Frenk} C.~S.,  {White} S.~D.~M.,  {Efstathiou} G.,   {Davis} M.,  1985,
  \mn@doi [\nat] {10.1038/317595a0}, \href
  {https://ui.adsabs.harvard.edu/abs/1985Natur.317..595F} {317, 595}

\bibitem[\protect\citeauthoryear{{Frenk}, {Evrard}, {White}  \&
  {Summers}}{{Frenk} et~al.}{1996}]{Frenk1996}
{Frenk} C.~S.,  {Evrard} A.~E.,  {White} S. D.~M.,   {Summers} F.~J.,  1996,
  \mn@doi [The Astrophysical Journal] {10.1086/178079}, \href
  {https://ui.adsabs.harvard.edu/abs/1996ApJ...472..460F} {472, 460}

\bibitem[\protect\citeauthoryear{{Gaia Collaboration} et~al.,}{{Gaia
  Collaboration} et~al.}{2018}]{Gaia2018}
{Gaia Collaboration} et~al., 2018, \mn@doi [\aap]
  {10.1051/0004-6361/201833051}, \href
  {https://ui.adsabs.harvard.edu/abs/2018A&A...616A...1G} {616, A1}

\bibitem[\protect\citeauthoryear{{Gao}, {White}, {Jenkins}, {Stoehr}  \&
  {Springel}}{{Gao} et~al.}{2004}]{Gao2004}
{Gao} L.,  {White} S.~D.~M.,  {Jenkins} A.,  {Stoehr} F.,   {Springel} V.,
  2004, \mn@doi [\mnras] {10.1111/j.1365-2966.2004.08360.x}, \href
  {http://adsabs.harvard.edu/abs/2004MNRAS.355..819G} {355, 819}

\bibitem[\protect\citeauthoryear{{Garrison-Kimmel}, {Bullock}, {Boylan-Kolchin}
   \& {Bardwell}}{{Garrison-Kimmel} et~al.}{2017}]{GarrisonKimmel2017}
{Garrison-Kimmel} S.,  {Bullock} J.~S.,  {Boylan-Kolchin} M.,   {Bardwell} E.,
  2017, \mn@doi [\mnras] {10.1093/mnras/stw2564}, \href
  {http://adsabs.harvard.edu/abs/2017MNRAS.464.3108G} {464, 3108}

\bibitem[\protect\citeauthoryear{{Garrison-Kimmel} et~al.,}{{Garrison-Kimmel}
  et~al.}{2019}]{GarrisonKimmel2019}
{Garrison-Kimmel} S.,  et~al., 2019, arXiv e-prints, \href
  {https://ui.adsabs.harvard.edu/abs/2019arXiv190310515G} {p. arXiv:1903.10515}

\bibitem[\protect\citeauthoryear{{Geha} et~al.,}{{Geha}
  et~al.}{2017}]{Geha2017}
{Geha} M.,  et~al., 2017, \mn@doi [\apj] {10.3847/1538-4357/aa8626}, \href
  {http://adsabs.harvard.edu/abs/2017ApJ...847....4G} {847, 4}

\bibitem[\protect\citeauthoryear{{Ghigna}, {Moore}, {Governato}, {Lake},
  {Quinn}  \& {Stadel}}{{Ghigna} et~al.}{2000}]{Ghigna2000}
{Ghigna} S.,  {Moore} B.,  {Governato} F.,  {Lake} G.,  {Quinn} T.,   {Stadel}
  J.,  2000, \mn@doi [\apj] {10.1086/317221}, \href
  {https://ui.adsabs.harvard.edu/abs/2000ApJ...544..616G} {544, 616}

\bibitem[\protect\citeauthoryear{{Gnedin}}{{Gnedin}}{2000}]{Gnedin2000}
{Gnedin} N.~Y.,  2000, \mn@doi [\apj] {10.1086/308876}, \href
  {http://adsabs.harvard.edu/abs/2000ApJ...535..530G} {535, 530}

\bibitem[\protect\citeauthoryear{{Gnedin} \& {Kaurov}}{{Gnedin} \&
  {Kaurov}}{2014}]{Gnedin2014}
{Gnedin} N.~Y.,  {Kaurov} A.~A.,  2014, \mn@doi [\apj]
  {10.1088/0004-637X/793/1/30}, \href
  {https://ui.adsabs.harvard.edu/abs/2014ApJ...793...30G} {793, 30}

\bibitem[\protect\citeauthoryear{{Grand} et~al.,}{{Grand}
  et~al.}{2017}]{Grand2017}
{Grand} R.~J.~J.,  et~al., 2017, \mn@doi [\mnras] {10.1093/mnras/stx071}, \href
  {http://adsabs.harvard.edu/abs/2017MNRAS.467..179G} {467, 179}

\bibitem[\protect\citeauthoryear{{Grand}, {Deason}, {White}, {Simpson},
  {G{\'o}mez}, {Marinacci}  \& {Pakmor}}{{Grand} et~al.}{2019}]{Grand2019}
{Grand} R. J.~J.,  {Deason} A.~J.,  {White} S. D.~M.,  {Simpson} C.~M.,
  {G{\'o}mez} F.~A.,  {Marinacci} F.,   {Pakmor} R.,  2019, \mn@doi [\mnras]
  {10.1093/mnrasl/slz092}, \href
  {https://ui.adsabs.harvard.edu/abs/2019MNRAS.487L..72G} {487, L72}

\bibitem[\protect\citeauthoryear{{Graus}, {Bullock}, {Kelley},
  {Boylan-Kolchin}, {Garrison-Kimmel}  \& {Qi}}{{Graus}
  et~al.}{2019}]{Graus2018}
{Graus} A.~S.,  {Bullock} J.~S.,  {Kelley} T.,  {Boylan-Kolchin} M.,
  {Garrison-Kimmel} S.,   {Qi} Y.,  2019, \mn@doi [\mnras]
  {10.1093/mnras/stz1992}, \href
  {https://ui.adsabs.harvard.edu/abs/2019MNRAS.488.4585G} {488, 4585}

\bibitem[\protect\citeauthoryear{{Guo} et~al.,}{{Guo} et~al.}{2011}]{Guo2011}
{Guo} Q.,  et~al., 2011, \mn@doi [\mnras] {10.1111/j.1365-2966.2010.18114.x},
  \href {http://adsabs.harvard.edu/abs/2011MNRAS.413..101G} {413, 101}

\bibitem[\protect\citeauthoryear{{Han}, {Cole}, {Frenk}  \& {Jing}}{{Han}
  et~al.}{2016}]{Han2016}
{Han} J.,  {Cole} S.,  {Frenk} C.~S.,   {Jing} Y.,  2016, \mn@doi [\mnras]
  {10.1093/mnras/stv2900}, \href
  {https://ui.adsabs.harvard.edu/abs/2016MNRAS.457.1208H} {457, 1208}

\bibitem[\protect\citeauthoryear{{Hargis}, {Willman}  \& {Peter}}{{Hargis}
  et~al.}{2014}]{Hargis2014}
{Hargis} J.~R.,  {Willman} B.,   {Peter} A.~H.~G.,  2014, \mn@doi [\apjl]
  {10.1088/2041-8205/795/1/L13}, \href
  {http://adsabs.harvard.edu/abs/2014ApJ...795L..13H} {795, L13}

\bibitem[\protect\citeauthoryear{{Hellwing}, {Frenk}, {Cautun}, {Bose},
  {Helly}, {Jenkins}, {Sawala}  \& {Cytowski}}{{Hellwing}
  et~al.}{2016}]{Hellwing2016}
{Hellwing} W.~A.,  {Frenk} C.~S.,  {Cautun} M.,  {Bose} S.,  {Helly} J.,
  {Jenkins} A.,  {Sawala} T.,   {Cytowski} M.,  2016, \mn@doi [\mnras]
  {10.1093/mnras/stw214}, \href
  {http://adsabs.harvard.edu/abs/2016MNRAS.457.3492H} {457, 3492}

\bibitem[\protect\citeauthoryear{{Helmi}, {Babusiaux}, {Koppelman}, {Massari},
  {Veljanoski}  \& {Brown}}{{Helmi} et~al.}{2018}]{Helmi2018}
{Helmi} A.,  {Babusiaux} C.,  {Koppelman} H.~H.,  {Massari} D.,  {Veljanoski}
  J.,   {Brown} A. G.~A.,  2018, \mn@doi [\nat] {10.1038/s41586-018-0625-x},
  \href {https://ui.adsabs.harvard.edu/abs/2018Natur.563...85H} {563, 85}

\bibitem[\protect\citeauthoryear{{Henriques}, {White}, {Thomas}, {Angulo},
  {Guo}, {Lemson}, {Springel}  \& {Overzier}}{{Henriques}
  et~al.}{2015}]{Henriques2015}
{Henriques} B.~M.~B.,  {White} S.~D.~M.,  {Thomas} P.~A.,  {Angulo} R.,  {Guo}
  Q.,  {Lemson} G.,  {Springel} V.,   {Overzier} R.,  2015, \mn@doi [\mnras]
  {10.1093/mnras/stv705}, \href
  {http://adsabs.harvard.edu/abs/2015MNRAS.451.2663H} {451, 2663}

\bibitem[\protect\citeauthoryear{{Hopkins}, {Kere{\v{s}}}, {O{\~n}orbe},
  {Faucher-Gigu{\`e}re}, {Quataert}, {Murray}  \& {Bullock}}{{Hopkins}
  et~al.}{2014}]{Hopkins2014}
{Hopkins} P.~F.,  {Kere{\v{s}}} D.,  {O{\~n}orbe} J.,  {Faucher-Gigu{\`e}re}
  C.-A.,  {Quataert} E.,  {Murray} N.,   {Bullock} J.~S.,  2014, \mn@doi
  [\mnras] {10.1093/mnras/stu1738}, \href
  {https://ui.adsabs.harvard.edu/abs/2014MNRAS.445..581H} {445, 581}

\bibitem[\protect\citeauthoryear{{Hopkins} et~al.,}{{Hopkins}
  et~al.}{2018}]{Hopkins2018}
{Hopkins} P.~F.,  et~al., 2018, \mn@doi [\mnras] {10.1093/mnras/sty1690}, \href
  {https://ui.adsabs.harvard.edu/abs/2018MNRAS.480..800H} {480, 800}

\bibitem[\protect\citeauthoryear{{Jahn}, {Sales}, {Wetzel}, {Boylan-Kolchin},
  {Chan}, {El-Badry}, {Lazar}  \& {Bullock}}{{Jahn} et~al.}{2019}]{Jahn2019}
{Jahn} E.~D.,  {Sales} L.~V.,  {Wetzel} A.,  {Boylan-Kolchin} M.,  {Chan}
  T.~K.,  {El-Badry} K.,  {Lazar} A.,   {Bullock} J.~S.,  2019, arXiv e-prints,
  \href {https://ui.adsabs.harvard.edu/abs/2019arXiv190702979J} {p.
  arXiv:1907.02979}

\bibitem[\protect\citeauthoryear{{Jethwa}, {Erkal}  \& {Belokurov}}{{Jethwa}
  et~al.}{2016}]{Jethwa2016}
{Jethwa} P.,  {Erkal} D.,   {Belokurov} V.,  2016, \mn@doi [\mnras]
  {10.1093/mnras/stw1343}, \href
  {https://ui.adsabs.harvard.edu/abs/2016MNRAS.461.2212J} {461, 2212}

\bibitem[\protect\citeauthoryear{{Jiang}, {Helly}, {Cole}  \& {Frenk}}{{Jiang}
  et~al.}{2014}]{Jiang2014}
{Jiang} L.,  {Helly} J.~C.,  {Cole} S.,   {Frenk} C.~S.,  2014, \mn@doi
  [\mnras] {10.1093/mnras/stu390}, \href
  {https://ui.adsabs.harvard.edu/abs/2014MNRAS.440.2115J} {440, 2115}

\bibitem[\protect\citeauthoryear{{Kallivayalil}, {van der Marel}  \&
  {Alcock}}{{Kallivayalil} et~al.}{2006}]{Kallivayalil2006}
{Kallivayalil} N.,  {van der Marel} R.~P.,   {Alcock} C.,  2006, \mn@doi [\apj]
  {10.1086/508014}, \href
  {https://ui.adsabs.harvard.edu/abs/2006ApJ...652.1213K} {652, 1213}

\bibitem[\protect\citeauthoryear{{Kauffmann}, {White}  \&
  {Guiderdoni}}{{Kauffmann} et~al.}{1993}]{Kauffmann1993}
{Kauffmann} G.,  {White} S.~D.~M.,   {Guiderdoni} B.,  1993, \mn@doi [\mnras]
  {10.1093/mnras/264.1.201}, \href
  {http://adsabs.harvard.edu/abs/1993MNRAS.264..201K} {264, 201}

\bibitem[\protect\citeauthoryear{{Kelley}, {Bullock}, {Garrison-Kimmel},
  {Boylan-Kolchin}, {Pawlowski}  \& {Graus}}{{Kelley}
  et~al.}{2018}]{Kelley2018}
{Kelley} T.,  {Bullock} J.~S.,  {Garrison-Kimmel} S.,  {Boylan-Kolchin} M.,
  {Pawlowski} M.~S.,   {Graus} A.~S.,  2018, arXiv e-prints, \href
  {https://ui.adsabs.harvard.edu/abs/2018arXiv181112413K} {p. arXiv:1811.12413}

\bibitem[\protect\citeauthoryear{{Kennedy}, {Frenk}, {Cole}  \&
  {Benson}}{{Kennedy} et~al.}{2014}]{Kennedy2014}
{Kennedy} R.,  {Frenk} C.,  {Cole} S.,   {Benson} A.,  2014, \mn@doi [\mnras]
  {10.1093/mnras/stu719}, \href
  {https://ui.adsabs.harvard.edu/abs/2014MNRAS.442.2487K} {442, 2487}

\bibitem[\protect\citeauthoryear{{Kim}, {Jerjen}, {Mackey}, {Da Costa}  \&
  {Milone}}{{Kim} et~al.}{2015}]{Kim2015}
{Kim} D.,  {Jerjen} H.,  {Mackey} D.,  {Da Costa} G.~S.,   {Milone} A.~P.,
  2015, \mn@doi [\apjl] {10.1088/2041-8205/804/2/L44}, \href
  {http://adsabs.harvard.edu/abs/2015ApJ...804L..44K} {804, L44}

\bibitem[\protect\citeauthoryear{{Kim}, {Peter}  \& {Hargis}}{{Kim}
  et~al.}{2018}]{Kim2018}
{Kim} S.~Y.,  {Peter} A. H.~G.,   {Hargis} J.~R.,  2018, \mn@doi [\prl]
  {10.1103/PhysRevLett.121.211302}, \href
  {https://ui.adsabs.harvard.edu/abs/2018PhRvL.121u1302K} {121, 211302}

\bibitem[\protect\citeauthoryear{{Komatsu} et~al.,}{{Komatsu}
  et~al.}{2011}]{Komatsu2011}
{Komatsu} E.,  et~al., 2011, \mn@doi [\apjs] {10.1088/0067-0049/192/2/18},
  \href {http://adsabs.harvard.edu/abs/2011ApJS..192...18K} {192, 18}

\bibitem[\protect\citeauthoryear{{Kondapally}, {Russell}, {Conselice}  \&
  {Penny}}{{Kondapally} et~al.}{2018}]{Kondapally2018}
{Kondapally} R.,  {Russell} G.~A.,  {Conselice} C.~J.,   {Penny} S.~J.,  2018,
  \mn@doi [\mnras] {10.1093/mnras/sty2333}, \href
  {https://ui.adsabs.harvard.edu/abs/2018MNRAS.481.1759K} {481, 1759}

\bibitem[\protect\citeauthoryear{{Koposov} et~al.,}{{Koposov}
  et~al.}{2008}]{Koposov2008}
{Koposov} S.,  et~al., 2008, \mn@doi [\apj] {10.1086/589911}, \href
  {http://adsabs.harvard.edu/abs/2008ApJ...686..279K} {686, 279}

\bibitem[\protect\citeauthoryear{{Koposov}, {Belokurov}, {Torrealba}  \&
  {Evans}}{{Koposov} et~al.}{2015}]{Koposov2015}
{Koposov} S.~E.,  {Belokurov} V.,  {Torrealba} G.,   {Evans} N.~W.,  2015,
  \mn@doi [\apj] {10.1088/0004-637X/805/2/130}, \href
  {http://adsabs.harvard.edu/abs/2015ApJ...805..130K} {805, 130}

\bibitem[\protect\citeauthoryear{{Lacey} et~al.,}{{Lacey}
  et~al.}{2016}]{Lacey2016}
{Lacey} C.~G.,  et~al., 2016, \mn@doi [\mnras] {10.1093/mnras/stw1888}, \href
  {http://adsabs.harvard.edu/abs/2016MNRAS.462.3854L} {462, 3854}

\bibitem[\protect\citeauthoryear{{Laevens} et~al.,}{{Laevens}
  et~al.}{2015}]{Laevens2015}
{Laevens} B.~P.~M.,  et~al., 2015, \mn@doi [\apjl]
  {10.1088/2041-8205/802/2/L18}, \href
  {http://adsabs.harvard.edu/abs/2015ApJ...802L..18L} {802, L18}

\bibitem[\protect\citeauthoryear{{Lagos}, {Baugh}, {Lacey}, {Benson}, {Kim}  \&
  {Power}}{{Lagos} et~al.}{2011}]{Lagos2011}
{Lagos} C.~D.~P.,  {Baugh} C.~M.,  {Lacey} C.~G.,  {Benson} A.~J.,  {Kim}
  H.-S.,   {Power} C.,  2011, \mn@doi [\mnras]
  {10.1111/j.1365-2966.2011.19583.x}, \href
  {http://adsabs.harvard.edu/abs/2011MNRAS.418.1649L} {418, 1649}

\bibitem[\protect\citeauthoryear{{Loeb} \& {Barkana}}{{Loeb} \&
  {Barkana}}{2001}]{Loeb2001}
{Loeb} A.,  {Barkana} R.,  2001, \mn@doi [\araa]
  {10.1146/annurev.astro.39.1.19}, \href
  {http://adsabs.harvard.edu/abs/2001ARA%26A..39...19L} {39, 19}

\bibitem[\protect\citeauthoryear{{Lovell} et~al.,}{{Lovell}
  et~al.}{2012}]{Lovell2012}
{Lovell} M.~R.,  et~al., 2012, \mn@doi [\mnras]
  {10.1111/j.1365-2966.2011.20200.x}, \href
  {https://ui.adsabs.harvard.edu/abs/2012MNRAS.420.2318L} {420, 2318}

\bibitem[\protect\citeauthoryear{{Lovell} et~al.,}{{Lovell}
  et~al.}{2016}]{Lovell2016}
{Lovell} M.~R.,  et~al., 2016, \mn@doi [\mnras] {10.1093/mnras/stw1317}, \href
  {https://ui.adsabs.harvard.edu/abs/2016MNRAS.461...60L} {461, 60}

\bibitem[\protect\citeauthoryear{{Ludlow}, {Navarro}, {Springel}, {Jenkins},
  {Frenk}  \& {Helmi}}{{Ludlow} et~al.}{2009}]{Ludlow2009}
{Ludlow} A.~D.,  {Navarro} J.~F.,  {Springel} V.,  {Jenkins} A.,  {Frenk}
  C.~S.,   {Helmi} A.,  2009, \mn@doi [\apj] {10.1088/0004-637X/692/1/931},
  \href {https://ui.adsabs.harvard.edu/abs/2009ApJ...692..931L} {692, 931}

\bibitem[\protect\citeauthoryear{{Macci{\`o}} \& {Fontanot}}{{Macci{\`o}} \&
  {Fontanot}}{2010}]{Maccio2010b}
{Macci{\`o}} A.~V.,  {Fontanot} F.,  2010, \mn@doi [\mnras]
  {10.1111/j.1745-3933.2010.00825.x}, \href
  {https://ui.adsabs.harvard.edu/abs/2010MNRAS.404L..16M} {404, L16}

\bibitem[\protect\citeauthoryear{{Macci{\`o}}, {Kang}, {Fontanot},
  {Somerville}, {Koposov}  \& {Monaco}}{{Macci{\`o}} et~al.}{2010}]{Maccio2010}
{Macci{\`o}} A.~V.,  {Kang} X.,  {Fontanot} F.,  {Somerville} R.~S.,  {Koposov}
  S.,   {Monaco} P.,  2010, \mn@doi [\mnras]
  {10.1111/j.1365-2966.2009.16031.x}, \href
  {https://ui.adsabs.harvard.edu/abs/2010MNRAS.402.1995M} {402, 1995}

\bibitem[\protect\citeauthoryear{{Mackereth} et~al.,}{{Mackereth}
  et~al.}{2019}]{Mackereth2019}
{Mackereth} J.~T.,  et~al., 2019, \mn@doi [\mnras] {10.1093/mnras/sty2955},
  \href {https://ui.adsabs.harvard.edu/abs/2019MNRAS.482.3426M} {482, 3426}

\bibitem[\protect\citeauthoryear{{Maraston}}{{Maraston}}{2005}]{Maraston2005}
{Maraston} C.,  2005, \mn@doi [\mnras] {10.1111/j.1365-2966.2005.09270.x},
  \href {http://adsabs.harvard.edu/abs/2005MNRAS.362..799M} {362, 799}

\bibitem[\protect\citeauthoryear{{McBride}, {Fakhouri}  \& {Ma}}{{McBride}
  et~al.}{2009}]{McBride2009}
{McBride} J.,  {Fakhouri} O.,   {Ma} C.-P.,  2009, \mn@doi [\mnras]
  {10.1111/j.1365-2966.2009.15329.x}, \href
  {https://ui.adsabs.harvard.edu/abs/2009MNRAS.398.1858M} {398, 1858}

\bibitem[\protect\citeauthoryear{{Munshi}, {Brooks}, {Christensen},
  {Applebaum}, {Holley-Bockelmann}, {Quinn}  \& {Wadsley}}{{Munshi}
  et~al.}{2019}]{Munshi2019}
{Munshi} F.,  {Brooks} A.~M.,  {Christensen} C.,  {Applebaum} E.,
  {Holley-Bockelmann} K.,  {Quinn} T.~R.,   {Wadsley} J.,  2019, \mn@doi [\apj]
  {10.3847/1538-4357/ab0085}, \href
  {https://ui.adsabs.harvard.edu/abs/2019ApJ...874...40M} {874, 40}

\bibitem[\protect\citeauthoryear{{Myeong}, {Evans}, {Belokurov}, {Sand ers}  \&
  {Koposov}}{{Myeong} et~al.}{2018}]{Myeong2018}
{Myeong} G.~C.,  {Evans} N.~W.,  {Belokurov} V.,  {Sand ers} J.~L.,   {Koposov}
  S.~E.,  2018, \mn@doi [\apjl] {10.3847/2041-8213/aad7f7}, \href
  {https://ui.adsabs.harvard.edu/abs/2018ApJ...863L..28M} {863, L28}

\bibitem[\protect\citeauthoryear{{Nagai} \& {Kravtsov}}{{Nagai} \&
  {Kravtsov}}{2005}]{Nagai2005}
{Nagai} D.,  {Kravtsov} A.~V.,  2005, \mn@doi [\apj] {10.1086/426016}, \href
  {https://ui.adsabs.harvard.edu/abs/2005ApJ...618..557N} {618, 557}

\bibitem[\protect\citeauthoryear{{Navarro}, {Eke}  \& {Frenk}}{{Navarro}
  et~al.}{1996}]{NEF96}
{Navarro} J.~F.,  {Eke} V.~R.,   {Frenk} C.~S.,  1996, \mn@doi [Monthly Notices
  of the Royal Astronomical Society] {10.1093/mnras/283.3.L72}, \href
  {https://ui.adsabs.harvard.edu/abs/1996MNRAS.283L..72N} {283, L72}

\bibitem[\protect\citeauthoryear{{Newton}, {Cautun}, {Jenkins}, {Frenk}  \&
  {Helly}}{{Newton} et~al.}{2018a}]{Newton2018}
{Newton} O.,  {Cautun} M.,  {Jenkins} A.,  {Frenk} C.~S.,   {Helly} J.~C.,
  2018a, arXiv e-prints, \href
  {https://ui.adsabs.harvard.edu/abs/2018arXiv180909625N} {p. arXiv:1809.09625}

\bibitem[\protect\citeauthoryear{{Newton}, {Cautun}, {Jenkins}, {Frenk}  \&
  {Helly}}{{Newton} et~al.}{2018b}]{Newton2017}
{Newton} O.,  {Cautun} M.,  {Jenkins} A.,  {Frenk} C.~S.,   {Helly} J.~C.,
  2018b, \mn@doi [\mnras] {10.1093/mnras/sty1085}, \href
  {https://ui.adsabs.harvard.edu/abs/2018MNRAS.479.2853N} {479, 2853}

\bibitem[\protect\citeauthoryear{{Okamoto}, {Gao}  \& {Theuns}}{{Okamoto}
  et~al.}{2008}]{Okamoto2008}
{Okamoto} T.,  {Gao} L.,   {Theuns} T.,  2008, \mn@doi [\mnras]
  {10.1111/j.1365-2966.2008.13830.x}, \href
  {http://adsabs.harvard.edu/abs/2008MNRAS.390..920O} {390, 920}

\bibitem[\protect\citeauthoryear{{Okamoto}, {Frenk}, {Jenkins}  \&
  {Theuns}}{{Okamoto} et~al.}{2010}]{Okamoto2010}
{Okamoto} T.,  {Frenk} C.~S.,  {Jenkins} A.,   {Theuns} T.,  2010, \mn@doi
  [Monthly Notices of the Royal Astronomical Society]
  {10.1111/j.1365-2966.2010.16690.x}, \href
  {https://ui.adsabs.harvard.edu/abs/2010MNRAS.406..208O} {406, 208}

\bibitem[\protect\citeauthoryear{{Patel}, {Besla}, {Mandel}  \& {Sohn}}{{Patel}
  et~al.}{2018}]{Patel2018}
{Patel} E.,  {Besla} G.,  {Mandel} K.,   {Sohn} S.~T.,  2018, \mn@doi [\apj]
  {10.3847/1538-4357/aab78f}, \href
  {https://ui.adsabs.harvard.edu/abs/2018ApJ...857...78P} {857, 78}

\bibitem[\protect\citeauthoryear{{Pe{\~n}arrubia}, {Benson}, {Walker},
  {Gilmore}, {McConnachie}  \& {Mayer}}{{Pe{\~n}arrubia}
  et~al.}{2010}]{Penarrubia2010}
{Pe{\~n}arrubia} J.,  {Benson} A.~J.,  {Walker} M.~G.,  {Gilmore} G.,
  {McConnachie} A.~W.,   {Mayer} L.,  2010, \mn@doi [\mnras]
  {10.1111/j.1365-2966.2010.16762.x}, \href
  {https://ui.adsabs.harvard.edu/abs/2010MNRAS.406.1290P} {406, 1290}

\bibitem[\protect\citeauthoryear{{Pe{\~n}arrubia}, {G{\'o}mez}, {Besla},
  {Erkal}  \& {Ma}}{{Pe{\~n}arrubia} et~al.}{2016}]{Penarrubia2016}
{Pe{\~n}arrubia} J.,  {G{\'o}mez} F.~A.,  {Besla} G.,  {Erkal} D.,   {Ma}
  Y.-Z.,  2016, \mn@doi [\mnras] {10.1093/mnrasl/slv160}, \href
  {https://ui.adsabs.harvard.edu/abs/2016MNRAS.456L..54P} {456, L54}

\bibitem[\protect\citeauthoryear{{Press} \& {Schechter}}{{Press} \&
  {Schechter}}{1974}]{Press1974}
{Press} W.~H.,  {Schechter} P.,  1974, \mn@doi [\apj] {10.1086/152650}, \href
  {https://ui.adsabs.harvard.edu/abs/1974ApJ...187..425P} {187, 425}

\bibitem[\protect\citeauthoryear{{Read}, {Walker}  \& {Steger}}{{Read}
  et~al.}{2019}]{Read2019}
{Read} J.~I.,  {Walker} M.~G.,   {Steger} P.,  2019, \mn@doi [\mnras]
  {10.1093/mnras/sty3404}, \href
  {https://ui.adsabs.harvard.edu/abs/2019MNRAS.484.1401R} {484, 1401}

\bibitem[\protect\citeauthoryear{{Reddick}, {Wechsler}, {Tinker}  \&
  {Behroozi}}{{Reddick} et~al.}{2013}]{Reddick2013}
{Reddick} R.~M.,  {Wechsler} R.~H.,  {Tinker} J.~L.,   {Behroozi} P.~S.,  2013,
  \mn@doi [\apj] {10.1088/0004-637X/771/1/30}, \href
  {http://adsabs.harvard.edu/abs/2013ApJ...771...30R} {771, 30}

\bibitem[\protect\citeauthoryear{{Rees}}{{Rees}}{1986}]{Rees1986}
{Rees} M.~J.,  1986, \mn@doi [\mnras] {10.1093/mnras/218.1.25P}, \href
  {http://adsabs.harvard.edu/abs/1986MNRAS.218P..25R} {218, 25P}

\bibitem[\protect\citeauthoryear{{Richings}, {Frenk}, {Jenkins}  \&
  {Robertson}}{{Richings} et~al.}{2018}]{Richings2018}
{Richings} J.,  {Frenk} C.,  {Jenkins} A.,   {Robertson} A.,  2018, arXiv
  e-prints, \href {https://ui.adsabs.harvard.edu/\#abs/2018arXiv181112437R} {p.
  arXiv:1811.12437}

\bibitem[\protect\citeauthoryear{{Sales}, {Navarro}, {Cooper}, {White}, {Frenk}
   \& {Helmi}}{{Sales} et~al.}{2011}]{Sales2011}
{Sales} L.~V.,  {Navarro} J.~F.,  {Cooper} A.~P.,  {White} S. D.~M.,  {Frenk}
  C.~S.,   {Helmi} A.,  2011, \mn@doi [\mnras]
  {10.1111/j.1365-2966.2011.19514.x}, \href
  {https://ui.adsabs.harvard.edu/abs/2011MNRAS.418..648S} {418, 648}

\bibitem[\protect\citeauthoryear{{Sales}, {Navarro}, {Kallivayalil}  \&
  {Frenk}}{{Sales} et~al.}{2017}]{Sales2017}
{Sales} L.~V.,  {Navarro} J.~F.,  {Kallivayalil} N.,   {Frenk} C.~S.,  2017,
  \mn@doi [\mnras] {10.1093/mnras/stw2816}, \href
  {https://ui.adsabs.harvard.edu/abs/2017MNRAS.465.1879S} {465, 1879}

\bibitem[\protect\citeauthoryear{{Samuel} et~al.,}{{Samuel}
  et~al.}{2019}]{Samuel2019}
{Samuel} J.,  et~al., 2019, arXiv e-prints, \href
  {https://ui.adsabs.harvard.edu/abs/2019arXiv190411508S} {p. arXiv:1904.11508}

\bibitem[\protect\citeauthoryear{{Sawala} et~al.,}{{Sawala}
  et~al.}{2016}]{Sawala2016}
{Sawala} T.,  et~al., 2016, \mn@doi [\mnras] {10.1093/mnras/stw145}, \href
  {http://adsabs.harvard.edu/abs/2016MNRAS.457.1931S} {457, 1931}

\bibitem[\protect\citeauthoryear{{Sawala}, {Pihajoki}, {Johansson}, {Frenk},
  {Navarro}, {Oman}  \& {White}}{{Sawala} et~al.}{2017}]{Sawala2017}
{Sawala} T.,  {Pihajoki} P.,  {Johansson} P.~H.,  {Frenk} C.~S.,  {Navarro}
  J.~F.,  {Oman} K.~A.,   {White} S.~D.~M.,  2017, \mn@doi [\mnras]
  {10.1093/mnras/stx360}, \href
  {http://adsabs.harvard.edu/abs/2017MNRAS.467.4383S} {467, 4383}

\bibitem[\protect\citeauthoryear{{Shao}, {Cautun}, {Deason}, {Frenk}  \&
  {Theuns}}{{Shao} et~al.}{2018}]{Shao2018}
{Shao} S.,  {Cautun} M.,  {Deason} A.~J.,  {Frenk} C.~S.,   {Theuns} T.,  2018,
  \mn@doi [\mnras] {10.1093/mnras/sty1470}, \href
  {https://ui.adsabs.harvard.edu/abs/2018MNRAS.479..284S} {479, 284}

\bibitem[\protect\citeauthoryear{{Sheth} \& {Tormen}}{{Sheth} \&
  {Tormen}}{2004}]{Sheth2004}
{Sheth} R.~K.,  {Tormen} G.,  2004, \mn@doi [\mnras]
  {10.1111/j.1365-2966.2004.07733.x}, \href
  {http://adsabs.harvard.edu/abs/2004MNRAS.350.1385S} {350, 1385}

\bibitem[\protect\citeauthoryear{{Simha} \& {Cole}}{{Simha} \&
  {Cole}}{2017}]{Simha2017}
{Simha} V.,  {Cole} S.,  2017, \mn@doi [\mnras] {10.1093/mnras/stx1942}, \href
  {http://adsabs.harvard.edu/abs/2017MNRAS.472.1392S} {472, 1392}

\bibitem[\protect\citeauthoryear{{Simpson}, {Bryan}, {Johnston}, {Smith}, {Mac
  Low}, {Sharma}  \& {Tumlinson}}{{Simpson} et~al.}{2013}]{Simpson2013}
{Simpson} C.~M.,  {Bryan} G.~L.,  {Johnston} K.~V.,  {Smith} B.~D.,  {Mac Low}
  M.-M.,  {Sharma} S.,   {Tumlinson} J.,  2013, \mn@doi [\mnras]
  {10.1093/mnras/stt474}, \href
  {https://ui.adsabs.harvard.edu/abs/2013MNRAS.432.1989S} {432, 1989}

\bibitem[\protect\citeauthoryear{{Simpson}, {Grand}, {G{\'o}mez}, {Marinacci},
  {Pakmor}, {Springel}, {Campbell}  \& {Frenk}}{{Simpson}
  et~al.}{2018}]{Simpson2018}
{Simpson} C.~M.,  {Grand} R. J.~J.,  {G{\'o}mez} F.~A.,  {Marinacci} F.,
  {Pakmor} R.,  {Springel} V.,  {Campbell} D. J.~R.,   {Frenk} C.~S.,  2018,
  \mn@doi [\mnras] {10.1093/mnras/sty774}, \href
  {https://ui.adsabs.harvard.edu/abs/2018MNRAS.478..548S} {478, 548}

\bibitem[\protect\citeauthoryear{{Smith} et~al.,}{{Smith}
  et~al.}{2007}]{Smith2007}
{Smith} M.~C.,  et~al., 2007, \mn@doi [\mnras]
  {10.1111/j.1365-2966.2007.11964.x}, \href
  {http://adsabs.harvard.edu/abs/2007MNRAS.379..755S} {379, 755}

\bibitem[\protect\citeauthoryear{{Somerville} \& {Primack}}{{Somerville} \&
  {Primack}}{1999}]{Somerville1999}
{Somerville} R.~S.,  {Primack} J.~R.,  1999, \mn@doi [\mnras]
  {10.1046/j.1365-8711.1999.03032.x}, \href
  {http://adsabs.harvard.edu/abs/1999MNRAS.310.1087S} {310, 1087}

\bibitem[\protect\citeauthoryear{{Springel}}{{Springel}}{2005}]{Springel2005}
{Springel} V.,  2005, \mn@doi [\mnras] {10.1111/j.1365-2966.2005.09655.x},
  \href {http://adsabs.harvard.edu/abs/2005MNRAS.364.1105S} {364, 1105}

\bibitem[\protect\citeauthoryear{{Springel}, {Yoshida}  \& {White}}{{Springel}
  et~al.}{2001a}]{Springel2001}
{Springel} V.,  {Yoshida} N.,   {White} S.~D.~M.,  2001a, \mn@doi [\na]
  {10.1016/S1384-1076(01)00042-2}, \href
  {http://adsabs.harvard.edu/abs/2001NewA....6...79S} {6, 79}

\bibitem[\protect\citeauthoryear{{Springel}, {White}, {Tormen}  \&
  {Kauffmann}}{{Springel} et~al.}{2001b}]{Springel2001b}
{Springel} V.,  {White} S.~D.~M.,  {Tormen} G.,   {Kauffmann} G.,  2001b,
  \mn@doi [\mnras] {10.1046/j.1365-8711.2001.04912.x}, \href
  {http://adsabs.harvard.edu/abs/2001MNRAS.328..726S} {328, 726}

\bibitem[\protect\citeauthoryear{{Springel} et~al.,}{{Springel}
  et~al.}{2008}]{Springel2008}
{Springel} V.,  et~al., 2008, \mn@doi [\mnras]
  {10.1111/j.1365-2966.2008.14066.x}, \href
  {http://adsabs.harvard.edu/abs/2008MNRAS.391.1685S} {391, 1685}

\bibitem[\protect\citeauthoryear{{Tollerud}, {Bullock}, {Strigari}  \&
  {Willman}}{{Tollerud} et~al.}{2008}]{Tollerud2008}
{Tollerud} E.~J.,  {Bullock} J.~S.,  {Strigari} L.~E.,   {Willman} B.,  2008,
  \mn@doi [\apj] {10.1086/592102}, \href
  {http://adsabs.harvard.edu/abs/2008ApJ...688..277T} {688, 277}

\bibitem[\protect\citeauthoryear{{Wang} et~al.,}{{Wang}
  et~al.}{2011}]{Wang2011}
{Wang} J.,  et~al., 2011, \mn@doi [\mnras] {10.1111/j.1365-2966.2011.18220.x},
  \href {https://ui.adsabs.harvard.edu/abs/2011MNRAS.413.1373W} {413, 1373}

\bibitem[\protect\citeauthoryear{{Wang}, {Frenk}, {Navarro}, {Gao}  \&
  {Sawala}}{{Wang} et~al.}{2012}]{Wang2012}
{Wang} J.,  {Frenk} C.~S.,  {Navarro} J.~F.,  {Gao} L.,   {Sawala} T.,  2012,
  \mn@doi [\mnras] {10.1111/j.1365-2966.2012.21357.x}, \href
  {https://ui.adsabs.harvard.edu/abs/2012MNRAS.424.2715W} {424, 2715}

\bibitem[\protect\citeauthoryear{{Wang}, {Han}, {Cooper}, {Cole}, {Frenk}  \&
  {Lowing}}{{Wang} et~al.}{2015}]{Wang2015}
{Wang} W.,  {Han} J.,  {Cooper} A.~P.,  {Cole} S.,  {Frenk} C.,   {Lowing} B.,
  2015, \mn@doi [\mnras] {10.1093/mnras/stv1647}, \href
  {http://adsabs.harvard.edu/abs/2015MNRAS.453..377W} {453, 377}

\bibitem[\protect\citeauthoryear{{Watkins}, {van der Marel}, {Sohn}  \&
  {Evans}}{{Watkins} et~al.}{2019}]{Watkins2019}
{Watkins} L.~L.,  {van der Marel} R.~P.,  {Sohn} S.~T.,   {Evans} N.~W.,  2019,
  \mn@doi [\apj] {10.3847/1538-4357/ab089f}, \href
  {https://ui.adsabs.harvard.edu/abs/2019ApJ...873..118W} {873, 118}

\bibitem[\protect\citeauthoryear{{Wechsler}, {Bullock}, {Primack}, {Kravtsov}
  \& {Dekel}}{{Wechsler} et~al.}{2002}]{Wechsler2002}
{Wechsler} R.~H.,  {Bullock} J.~S.,  {Primack} J.~R.,  {Kravtsov} A.~V.,
  {Dekel} A.,  2002, \mn@doi [\apj] {10.1086/338765}, \href
  {https://ui.adsabs.harvard.edu/abs/2002ApJ...568...52W} {568, 52}

\bibitem[\protect\citeauthoryear{{Weisz}, {Dolphin}, {Skillman}, {Holtzman},
  {Gilbert}, {Dalcanton}  \& {Williams}}{{Weisz} et~al.}{2014}]{Weisz2014}
{Weisz} D.~R.,  {Dolphin} A.~E.,  {Skillman} E.~D.,  {Holtzman} J.,  {Gilbert}
  K.~M.,  {Dalcanton} J.~J.,   {Williams} B.~F.,  2014, \mn@doi [\apj]
  {10.1088/0004-637X/789/2/148}, \href
  {https://ui.adsabs.harvard.edu/abs/2014ApJ...789..148W} {789, 148}

\bibitem[\protect\citeauthoryear{{Wetzel}, {Hopkins}, {Kim},
  {Faucher-Gigu{\`e}re}, {Kere{\v{s}}}  \& {Quataert}}{{Wetzel}
  et~al.}{2016}]{Wetzel2016}
{Wetzel} A.~R.,  {Hopkins} P.~F.,  {Kim} J.-h.,  {Faucher-Gigu{\`e}re} C.-A.,
  {Kere{\v{s}}} D.,   {Quataert} E.,  2016, \mn@doi [\apjl]
  {10.3847/2041-8205/827/2/L23}, \href
  {https://ui.adsabs.harvard.edu/abs/2016ApJ...827L..23W} {827, L23}

\bibitem[\protect\citeauthoryear{{Wheeler}, {O{\~n}orbe}, {Bullock},
  {Boylan-Kolchin}, {Elbert}, {Garrison-Kimmel}, {Hopkins}  \&
  {Kere{\v{s}}}}{{Wheeler} et~al.}{2015}]{Wheeler2015}
{Wheeler} C.,  {O{\~n}orbe} J.,  {Bullock} J.~S.,  {Boylan-Kolchin} M.,
  {Elbert} O.~D.,  {Garrison-Kimmel} S.,  {Hopkins} P.~F.,   {Kere{\v{s}}} D.,
  2015, \mn@doi [\mnras] {10.1093/mnras/stv1691}, \href
  {https://ui.adsabs.harvard.edu/abs/2015MNRAS.453.1305W} {453, 1305}

\bibitem[\protect\citeauthoryear{{Wheeler} et~al.,}{{Wheeler}
  et~al.}{2018}]{Wheeler2018}
{Wheeler} C.,  et~al., 2018, arXiv e-prints, \href
  {https://ui.adsabs.harvard.edu/abs/2018arXiv181202749W} {p. arXiv:1812.02749}

\bibitem[\protect\citeauthoryear{{Yurin} \& {Springel}}{{Yurin} \&
  {Springel}}{2015}]{Yurin2015}
{Yurin} D.,  {Springel} V.,  2015, \mn@doi [\mnras] {10.1093/mnras/stv1454},
  \href {http://adsabs.harvard.edu/abs/2015MNRAS.452.2367Y} {452, 2367}

\bibitem[\protect\citeauthoryear{{Zaritsky}, {Smith}, {Frenk}  \&
  {White}}{{Zaritsky} et~al.}{1993}]{Zaritsky1993}
{Zaritsky} D.,  {Smith} R.,  {Frenk} C.,   {White} S. D.~M.,  1993, \mn@doi
  [\apj] {10.1086/172379}, \href
  {https://ui.adsabs.harvard.edu/abs/1993ApJ...405..464Z} {405, 464}

\bibitem[\protect\citeauthoryear{{Zehavi}, {Contreras}, {Padilla}, {Smith},
  {Baugh}  \& {Norberg}}{{Zehavi} et~al.}{2018}]{Zehavi2018}
{Zehavi} I.,  {Contreras} S.,  {Padilla} N.,  {Smith} N.~J.,  {Baugh} C.~M.,
  {Norberg} P.,  2018, \mn@doi [\apj] {10.3847/1538-4357/aaa54a}, \href
  {http://adsabs.harvard.edu/abs/2018ApJ...853...84Z} {853, 84}

\bibitem[\protect\citeauthoryear{{Zentner}, {Berlind}, {Bullock}, {Kravtsov}
  \& {Wechsler}}{{Zentner} et~al.}{2005}]{Zentner2005}
{Zentner} A.~R.,  {Berlind} A.~A.,  {Bullock} J.~S.,  {Kravtsov} A.~V.,
  {Wechsler} R.~H.,  2005, \mn@doi [\apj] {10.1086/428898}, \href
  {https://ui.adsabs.harvard.edu/\#abs/2005ApJ...624..505Z} {624, 505}

\bibitem[\protect\citeauthoryear{{Zhu}, {Marinacci}, {Maji}, {Li}, {Springel}
  \& {Hernquist}}{{Zhu} et~al.}{2016}]{Zhu2016}
{Zhu} Q.,  {Marinacci} F.,  {Maji} M.,  {Li} Y.,  {Springel} V.,   {Hernquist}
  L.,  2016, \mn@doi [\mnras] {10.1093/mnras/stw374}, \href
  {http://adsabs.harvard.edu/abs/2016MNRAS.458.1559Z} {458, 1559}

\bibitem[\protect\citeauthoryear{{van den Bosch} \& {Ogiya}}{{van den Bosch} \&
  {Ogiya}}{2018}]{vdB2018b}
{van den Bosch} F.~C.,  {Ogiya} G.,  2018, \mn@doi [\mnras]
  {10.1093/mnras/sty084}, \href
  {https://ui.adsabs.harvard.edu/abs/2018MNRAS.475.4066V} {475, 4066}

\bibitem[\protect\citeauthoryear{{van den Bosch}, {Yang}, {Mo}  \&
  {Norberg}}{{van den Bosch} et~al.}{2005}]{vdB2005}
{van den Bosch} F.~C.,  {Yang} X.,  {Mo} H.~J.,   {Norberg} P.,  2005, \mn@doi
  [\mnras] {10.1111/j.1365-2966.2004.08407.x}, \href
  {https://ui.adsabs.harvard.edu/abs/2005MNRAS.356.1233V} {356, 1233}

\bibitem[\protect\citeauthoryear{{van den Bosch}, {Ogiya}, {Hahn}  \&
  {Burkert}}{{van den Bosch} et~al.}{2018}]{vdB2018a}
{van den Bosch} F.~C.,  {Ogiya} G.,  {Hahn} O.,   {Burkert} A.,  2018, \mn@doi
  [\mnras] {10.1093/mnras/stx2956}, \href
  {https://ui.adsabs.harvard.edu/abs/2018MNRAS.474.3043V} {474, 3043}

\makeatother
\end{thebibliography}

%%%%%%%%%%%%%%%%%%%%%%%%%%%%%%%%%%%%%%%%%%%%%%%%%%

%%%%%%%%%%%%%%%%% APPENDICES %%%%%%%%%%%%%%%%%%%%%

\appendix

\section{disc disruption in hydrodynamical simulations}
\label{sect:disc_disrupt_sims}

In Section~\ref{sect:disc}, we saw that the destruction of satellite galaxies by the central galaxy reduces the size of the population substantially, particularly in the vicinity of the central disc. In this Appendix, we compare the extent of this depletion in the {\sc Fire-2}, {\sc Apostle} and {\sc Auriga} simulations.

\begin{figure}
    \centering
    \includegraphics[width=\columnwidth]{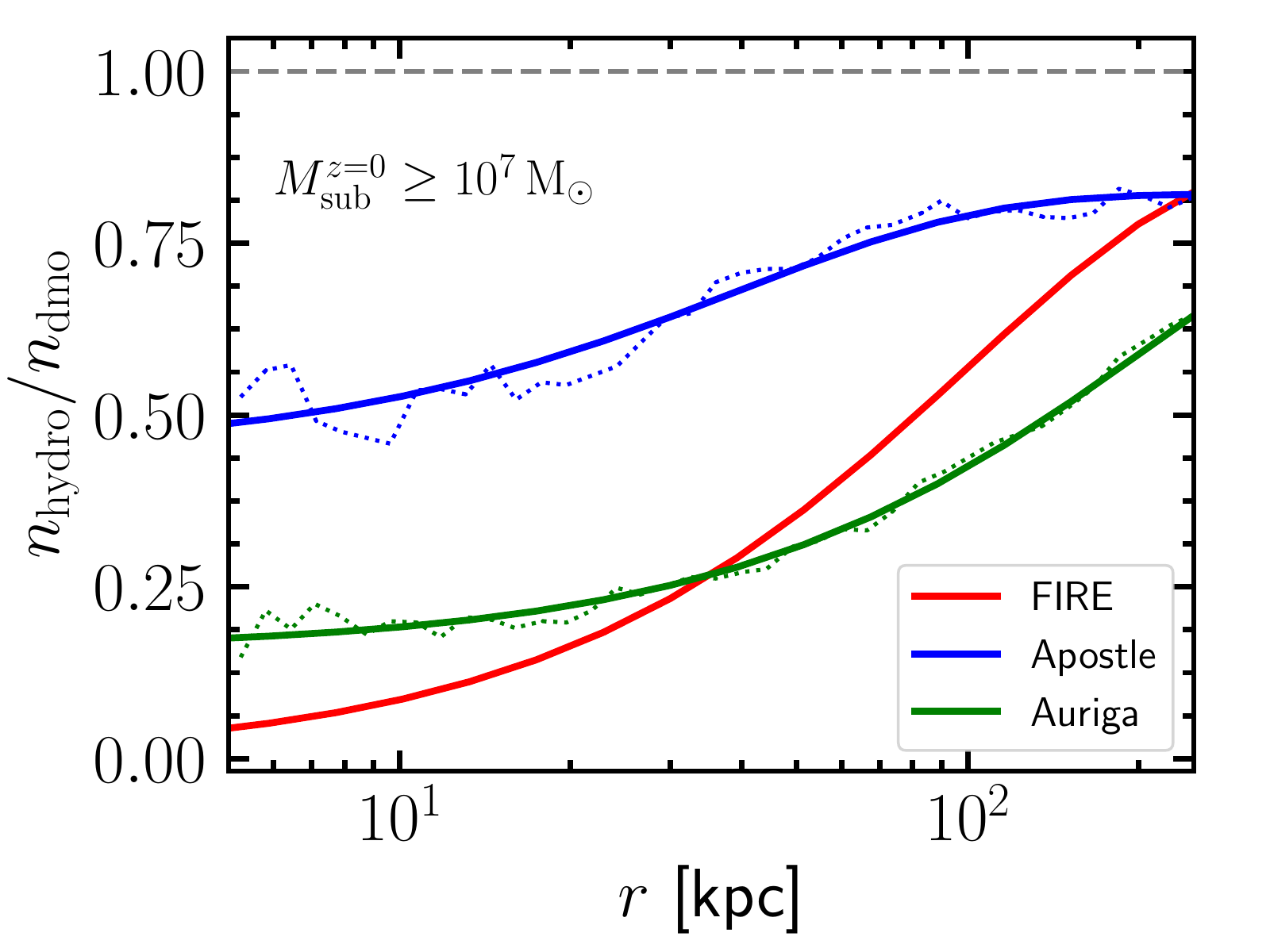}
    \caption{Ratio of the radial number density of resolved subhaloes with mass $M_{{\rm sub}}^{z=0}\geq 10^7\,{\rm M}_\odot$ in hydrodynamical and dark matter-only versions of the {\sc Fire-2}, {\sc Apostle} and {\sc Auriga} simulations. The thin dotted lines show the median ratios measured by \citet{Richings2018}, while the corresponding solid curves are fits to these ratios using the functional form suggested by \citet[][Eq.~\ref{eq:depletion}]{Samuel2019}. For each of these models, we list the best-fit values of the free parameters $\alpha$, $d_0$ and $d_1$ in Table~\ref{tab:radial_fit_params}.}
    \label{fig:hydro_radial_corr}
\end{figure}

Fig.~\ref{fig:hydro_radial_corr} shows the reduction in the number density of subhaloes as a function of radius in hydrodynamical versions of each simulation, compared to their dark matter-only counterparts. In all three cases, there is a general trend for the depletion to become more pronounced near the centre of the halo \citep[see also][]{Errani2017}. This is to be expected as the tidal forces experienced by subhaloes are largest in the vicinity of the central galaxy. The extent of the depletion at any given radius, however, is markedly different in each of the three simulations.

\begin{table}
    \centering
    \begin{tabular}{c|c|c|c|c}
    \hline
      Model & $\alpha$ & $d_0$ & $d_1$ & Reference \\
              &  & [kpc] & [kpc] \\
             \hline \hline
    {\sc Fire-2} & 0.9 (0.8) & 0 (8) & 100 (78) & \cite{Samuel2019} \\
              
    {\sc Apostle}  & 0.82 & 0 & 40 & \cite{Richings2018} \\
    {\sc Auriga}  & 0.85 & 0 & 207 & \cite{Richings2018} \\
    \hline
    \end{tabular}
    \caption{Best-fit values for the parameters describing the disc destruction correction function, $f(d)$ Eq.~\ref{eq:depletion}), in each of the models listed in the first column. For the {\sc Fire-2} model, the values in  parentheses correspond to satellites more massive than $M_\star\geq10^5\,{\rm M}_\odot$, as given by \citet{Samuel2019}.}
    \label{tab:radial_fit_params} 
\end{table}

It is clear that the degree of subhalo depletion is least severe in {\sc Apostle}. \cite{Richings2018} ascribe this to the fact that galactic discs in the highest resolution {\sc Apostle} simulations are 2 to 3 times less massive than the Milky Way disc.  In {\sc Auriga}, on the other hand, the central galaxy mass is perhaps too large by a factor of 1.5 to 2. The presence of more massive central galaxies in {\sc Auriga} and {\sc Fire-2} compared to {\sc Apostle}, causes subhaloes to experience stronger tidal forces during pericentric passages and thus to be destroyed more easily. This disruption extends well beyond the virial radius of the halo, and is particularly strong for subhaloes on radial orbits. \cite{Richings2018} provides a detailed explanation of the radial dependence of the depletion in the {\sc Apostle} and {\sc Auriga} simulations.

Another interesting feature of Fig.~\ref{fig:hydro_radial_corr} is that while the depletion near the halo centre is stronger in {\sc Fire-2} than in {\sc Auriga}, the trend reverses in the outer halo ($r>30\,$kpc). The difference in the outer parts is perhaps due to the fact that the central galaxies in {\sc Auriga} are more massive than those in {\sc Fire-2}. However, at $r<30\,$kpc, the depletion is less in {\sc Auriga} despite its more massive central galaxies. This may be due, at least in part, to the specific way in which the subhalo orbits are interpolated between snapshots in the two simulations. Due to the rarity of satellites in this region an accurate interpolation is crucial for estimating the average depletion, particularly near the halo centre. We refer the reader to \cite{Richings2018} for a comprehensive treatment of this topic.

\section{The importance of orphan galaxies}
\label{sect:ImpOrphan}

\begin{figure*}
    \centering
    \includegraphics[width=0.475\textwidth]{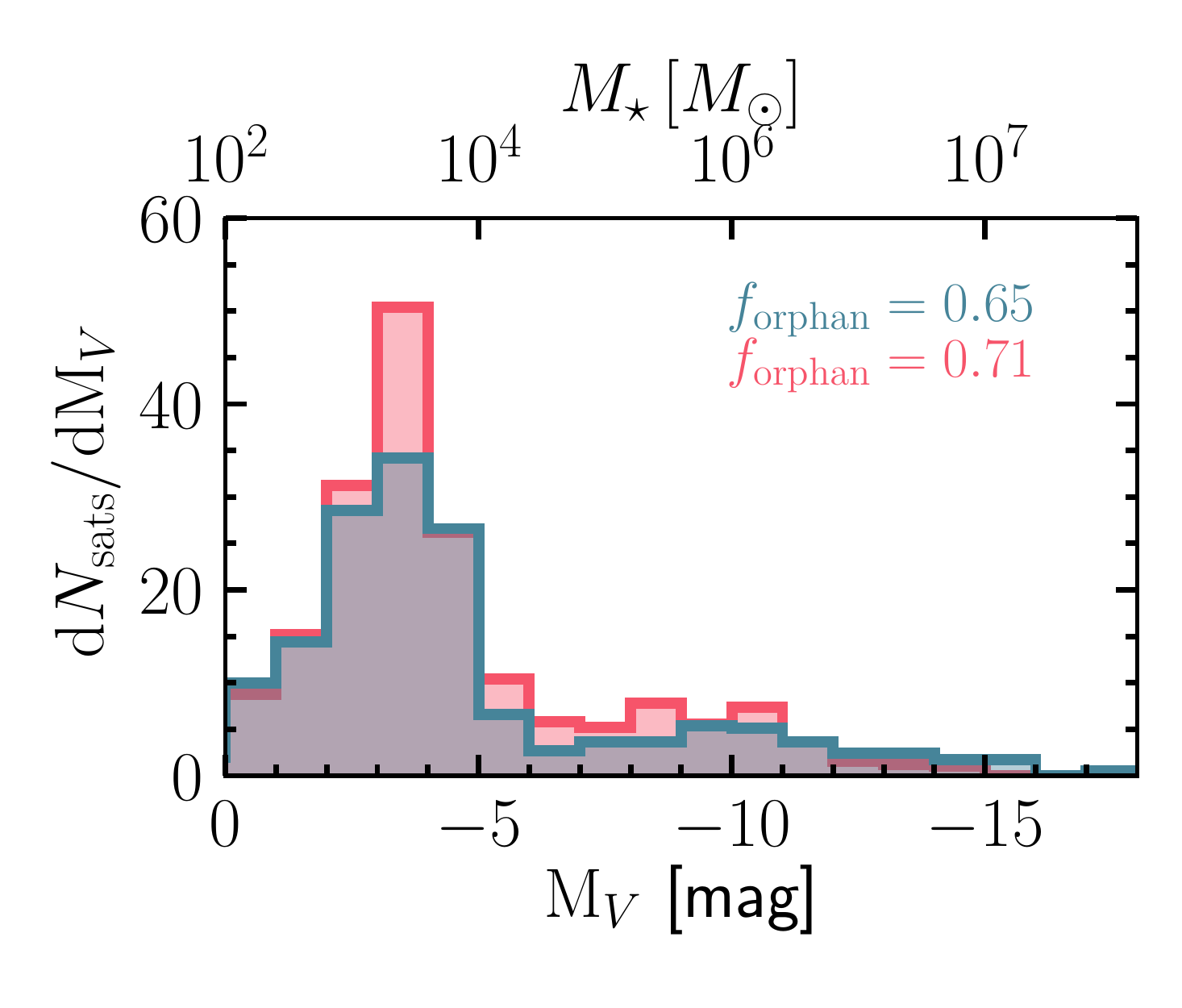}
    \includegraphics[width=0.475\textwidth]{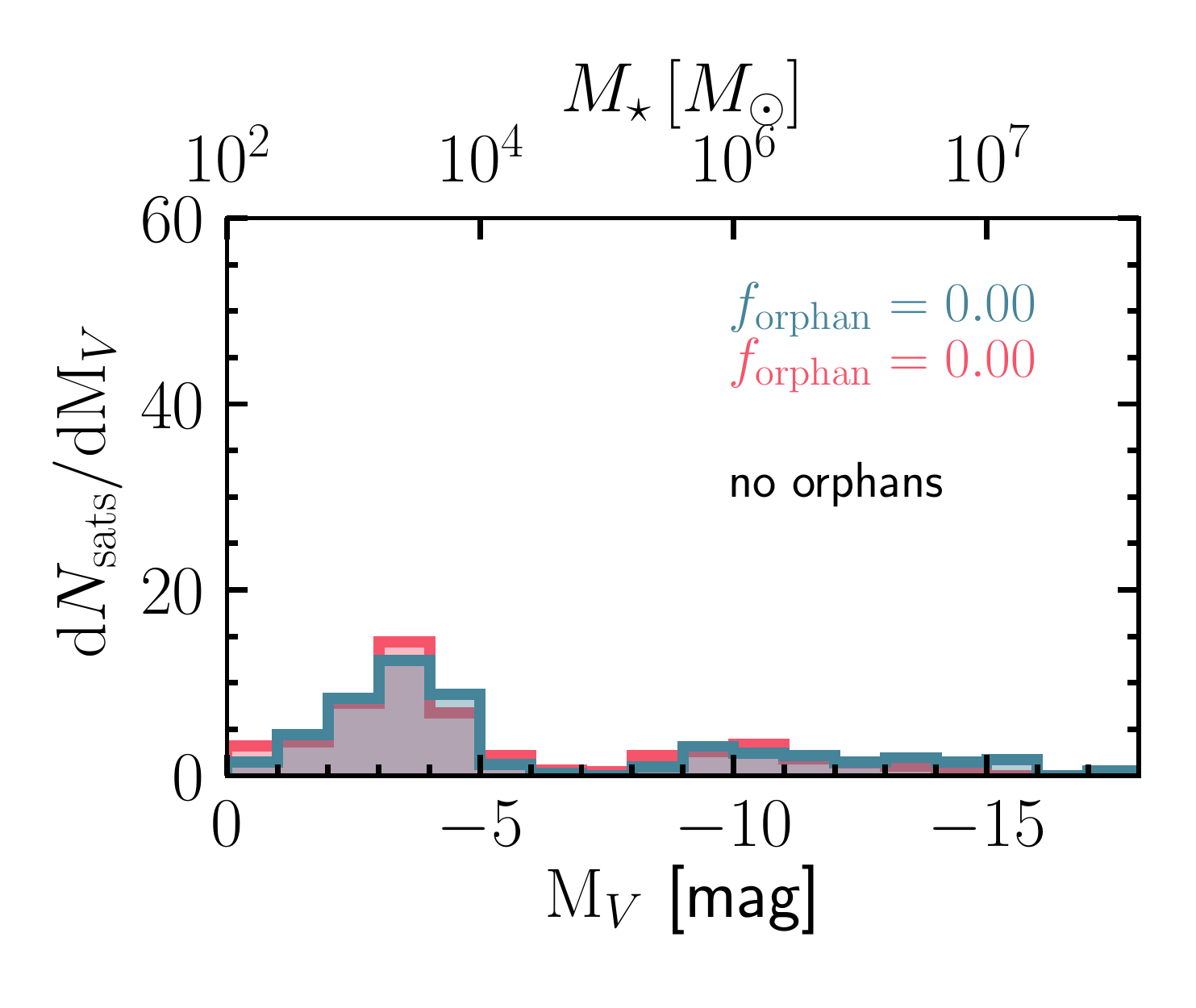}
    \caption{As Fig.~\ref{fig:mah_vs_uf}, but now showing the comparison between the luminosity functions with (left) and without (right) orphans. The fraction of ultrafaint satellites identified today as orphans is given in the legend. The fraction is large, 60-70\%; these galaxies would be missed entirely in current hydrodynamical simulations but can be followed in our semi-analytic model coupled to a very high resolution $N$-body simulation. When the orphan population is neglected, the luminosity functions are nearly identical for early (red) and late-forming (blue) haloes. The host halo mass range in this and subsequent figures is $M_{200} = 1-1.3\times10^{12}\,{\rm M}_\odot$}
    \label{fig:orphan_compare_lf_assembly}
\end{figure*}

\begin{figure*}
    \centering
    \includegraphics[width=0.475\textwidth]{Figures/RadDistr_COLOR_z0p50.pdf}
    \includegraphics[width=0.475\textwidth]{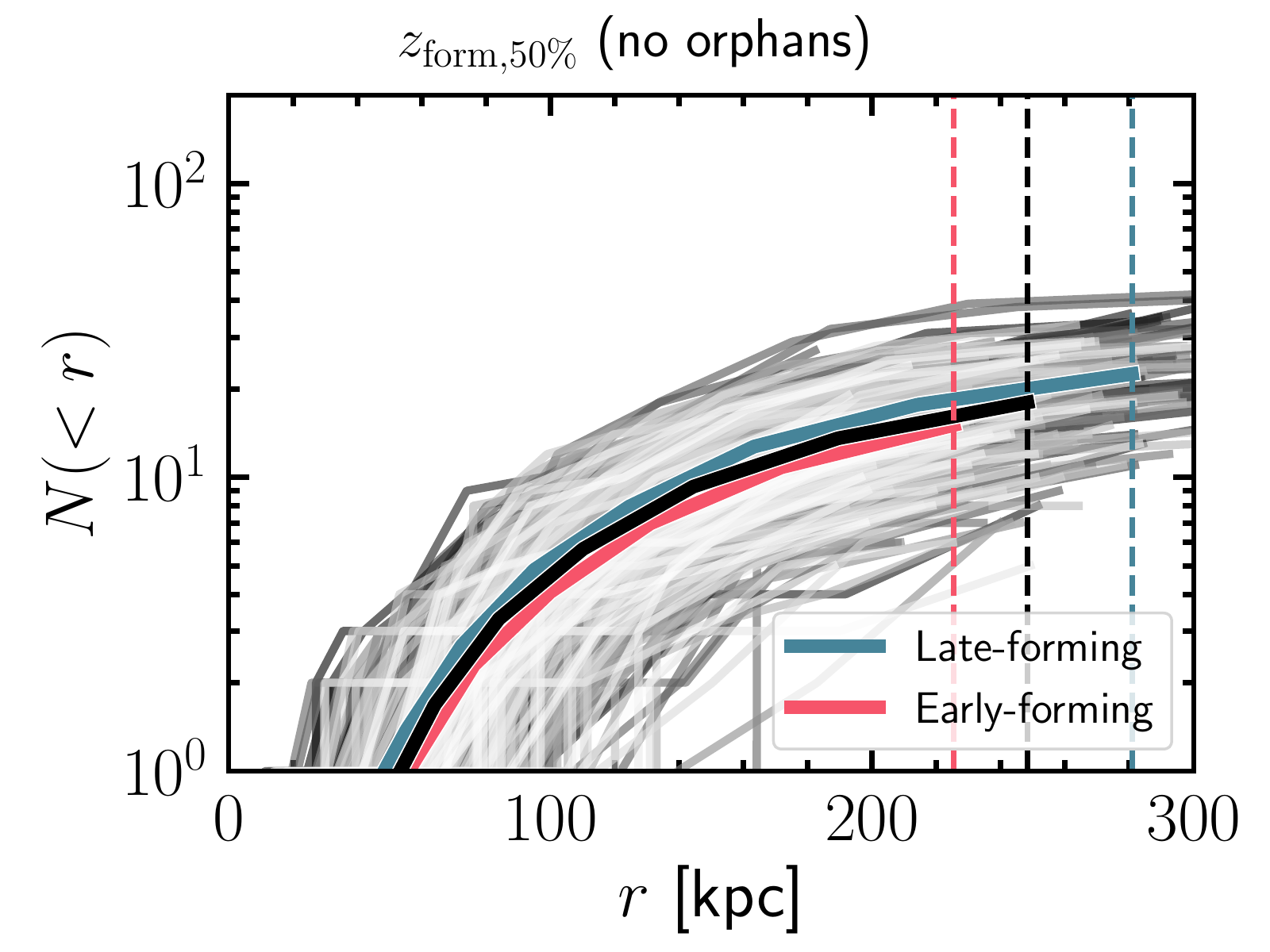}
    \caption{As Fig.~\ref{fig:radProf}, but now showing the comparison between radial profiles with (left) and without (right) orphans galaxies. Orphan galaxies make up a significant fraction of the total population at all radii and are by far the dominant population in the innermost 50 kpc.}
    \label{fig:orphans_compare_radial_profile} 
\end{figure*}

\begin{figure*}
    \centering
    \includegraphics[width=0.475\textwidth]{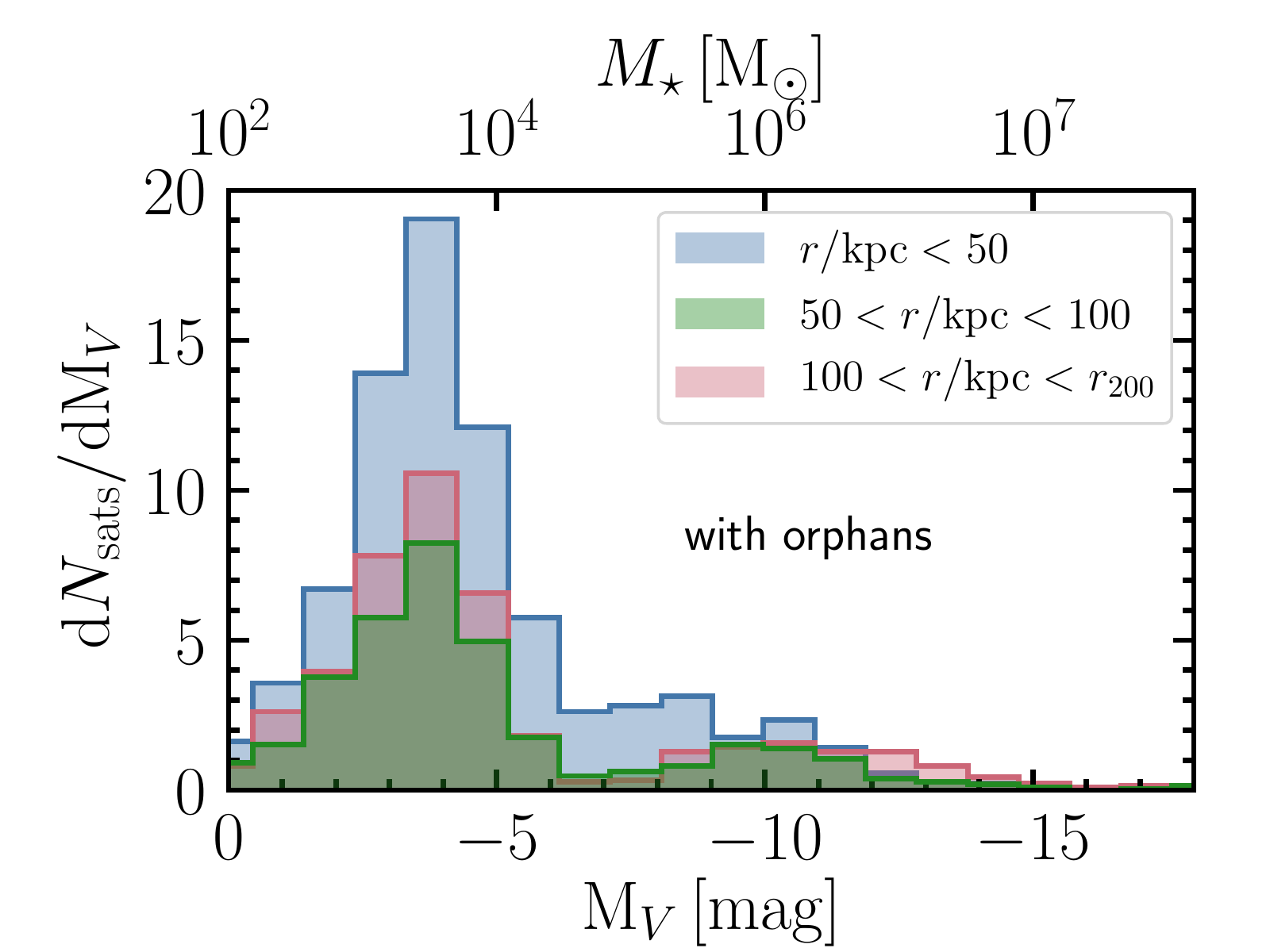}
    \includegraphics[width=0.475\textwidth]{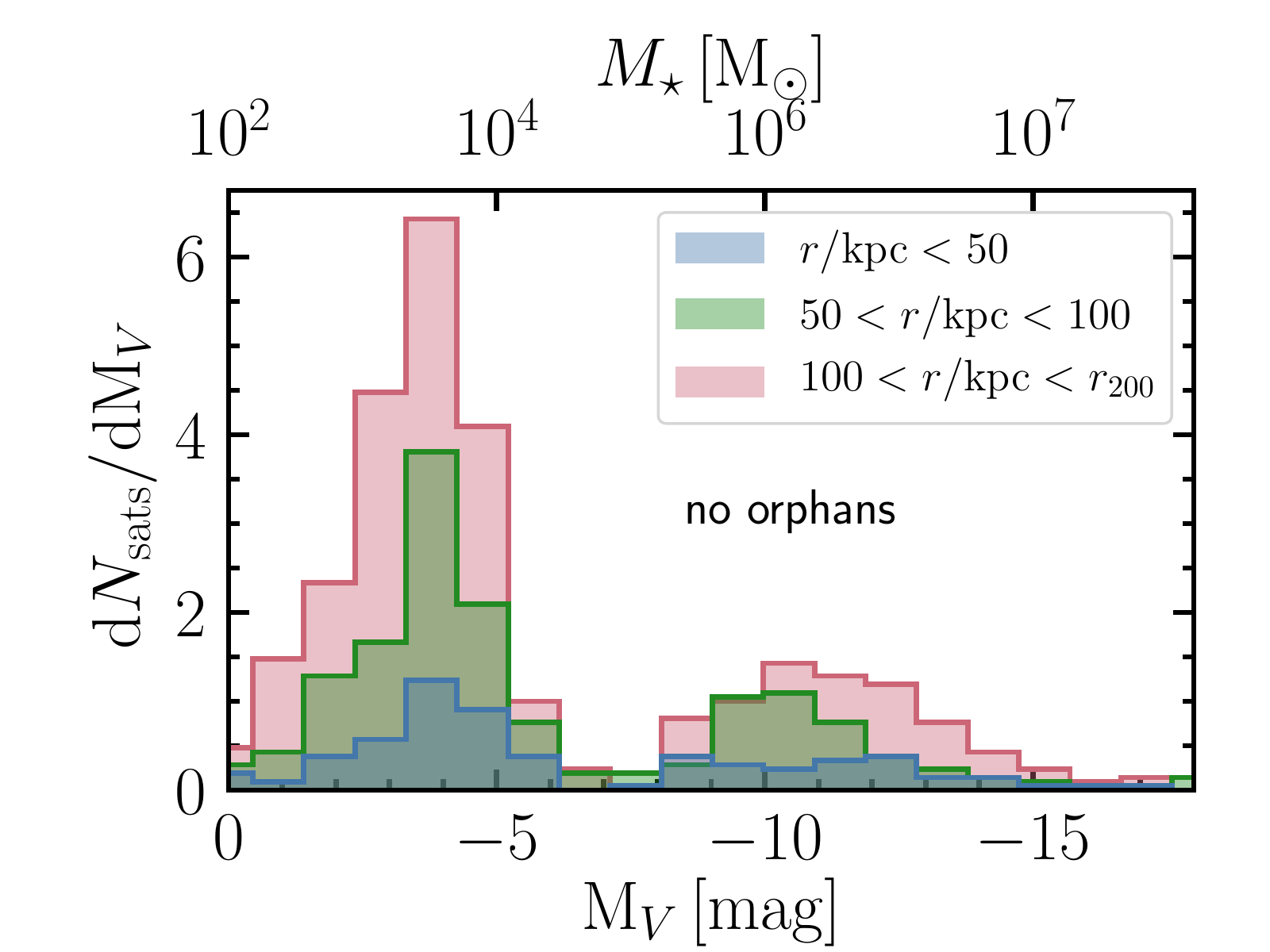}
    \caption{As Fig.~\ref{fig:lf_shells}, showing the luminosity
      function of satellites split into radial shells, with (left) and without (right) the inclusion of orphan galaxies (note the different $y$-axis limits in the two panels). The orphan satellites are distributed throughout the halo, with the majority of their contribution at the centre. Neglecting orphans results in a near absence of galaxies within the innermost 50 kpc.} 
    \label{fig:orphans_compare_lf_shells}
\end{figure*}

A key aspect of our analysis is the consistent tracking of ``orphan'' galaxies, a technique used to follow the dynamical evolution of \galform{} galaxies even after the dark matter subhalo in which they are hosted is disrupted below the nominal resolution limit of the \coco{} simulation ($\sim 3\times10^6 {\rm M}_\odot$ in halo mass; see Section~\ref{sect:Semia} for details). The results presented in the main body of the paper always include the orphan population; the purpose of this Appendix is to demonstrate how ignoring this population can lead to qualitatively different results.

Fig.~\ref{fig:orphan_compare_lf_assembly} revisits the connection between the number of ultrafaint satellites and the assembly history of the host halo, this time comparing the results with (left) and without (middle) orphan galaxies. The fraction of the total satellite population that is identified as orphans is given in each panel. This number, which is made up predominantly of ultrafaint satellites, is sizeable: around 65\% in the later-forming haloes, rising to over 70\% in early-forming haloes. Qualitatively, this dependence on halo assembly is expected. Early-forming haloes are likely to accrete satellites earlier than their late-forming counterparts. These satellites orbit and lose mass through tidal stripping, in many cases bringing their subhaloes below the \Subfind{} detection limit, resulting in the galaxies being tagged as orphans. On the other hand, satellites which fall into their hosts later experience fewer pericentric passages and thus a smaller fraction of them are likely to be tagged as orphans. Finally, the right panel in Fig.~\ref{fig:orphan_compare_lf_assembly} shows that neglecting the orphan population results in little to no difference in the satellite abundance in early-forming {\em vs} late-forming haloes.

Fig.~\ref{fig:orphans_compare_radial_profile} shows that there are also significant differences in the radial profiles of satellite galaxies when orphans are not included. While the total number of galaxies within a fixed radius, $r$, is, of course, significantly lower in all cases, there is also a change in the relative behaviour of early and late-forming haloes when ignoring orphan satellites. Indeed, the trends are inverted: at all $r$, late-forming haloes are now predicted to host more satellites than early-forming haloes, reproducing a common (but incorrect) conclusion often found in previous literature (see the discussion in Section~\ref{sect:assembly_number}), which is contrary to the result presented in Fig.~\ref{fig:radProf} (reproduced in the left-hand panel of Fig.~\ref{fig:orphans_compare_radial_profile}).

The radial profiles predicted by our semi-analytic model further highlight the importance of tracking orphan galaxies. These objects, which are largely ultrafaint galaxies, are distributed differently within the host halo relative to their more massive counterparts. The different spatial distributions are clearly illustrated in Fig.~\ref{fig:orphans_compare_lf_shells}, which shows the satellite luminosity function from the high resolution \coco{} simulation, split into radial shells. When orphans are not included (right panel), there is a near absence of satellites of any mass within the inner 50~kpc of the halo centre. Beyond this radius, we retrieve the familiar bimodal satellite distribution discussed in Section~\ref{sect:Semia}. In \coco{}, we find, on average, only around 6 satellites (across all masses) within the innermost 50 kpc when ignoring orphans; but when orphans are counted, there are on average $\sim$70 satellites within this same radius. As shown in Fig.~\ref{fig:orphans_compare_lf_shells}, the inner population is primarily made up of satellites with $M_\star\leq10^6\,{\rm M}_\odot$, although there is a small contribution from more massive satellites as well. Galaxies near the centres of haloes are more likely to be identified as orphans where tidal forces are more severe, and more likely to strip subhaloes below the resolution limit of the simulation. Furthermore, substructure finding algorithms often `lose' subhaloes near halo centres where the relatively low density contrast complicates the identification of substructures.

%%%%%%%%%%%%%%%%%%%%%%%%%%%%%%%%%%%%%%%%%%%%%%%%%%

% Don't change these lines
\bsp	% typesetting comment
\label{lastpage}
\end{document}